# Internet of Things Device Capabilities, Architectures, Protocols, and Smart Applications in Healthcare Domain: A Review

Md. Milon Islam, Sheikh Nooruddin, Fakhri Karray, *Fellow, IEEE*, and Ghulam Muhammad, *Senior Member, IEEE*

*Abstract*—Nowadays, the Internet has spread to practically every country around the world and is having unprecedented effects on people's lives. The Internet of Things (IoT) is getting more popular and has a high level of interest in both practitioners and academicians in the age of wireless communication due to its diverse applications. The IoT is a technology that enables everyday things to become savvier, everyday computation towards becoming intellectual, and everyday communication to become a little more insightful. In this paper, the most common and popular IoT device capabilities, architectures, and protocols are demonstrated in brief to provide a clear overview of the IoT technology to the researchers in this area. The common IoT device capabilities including hardware (Raspberry Pi, Arduino, and ESP8266) and software (operating systems, and built-in tools) platforms are described in detail. The widely used architectures that have been recently evolved and used are the three-layer architecture, SOA-based architecture, and middleware-based architecture. The popular protocols for IoT are demonstrated which include CoAP, MQTT, XMPP, AMQP, DDS, LoWPAN, BLE, and Zigbee that are frequently utilized to develop smart IoT applications. Additionally, this research provides an in-depth overview of the potential healthcare applications based on IoT technologies in the context of addressing various healthcare concerns. Finally, this paper summarizes state-of-the-art knowledge, highlights open issues and shortcomings, and provides recommendations for further studies which would be quite beneficial to anyone with a desire to work in this field and make breakthroughs to get expertise in this area.

*Index Terms*— Internet of Things, Device Capabilities, IoT Architecture, Communication Protocol, Healthcare Applications.

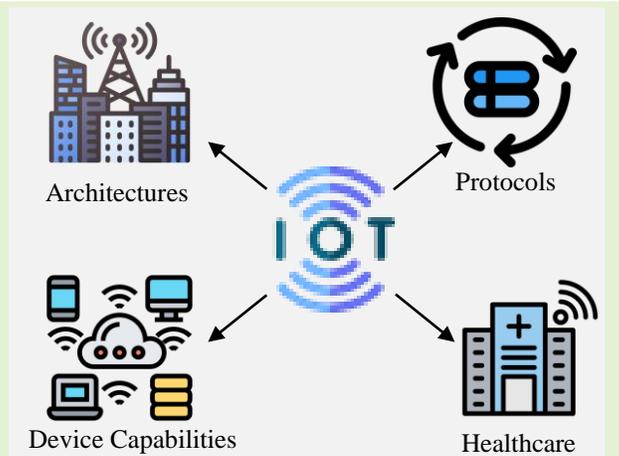

Architectures

Protocols

Device Capabilities

Healthcare

## I. Introduction

THE IoT concept represents a set of Internet-connected devices that are empowered with computing capability, are easily recognizable, and can share data over the Internet without the need for physical contact. The IoT revolution has resulted in unprecedented levels of interconnection among devices, allowing for the development of smart environments. The smart connectivity with the existing network, as well as

Md. Milon Islam, and Sheikh Nooruddin are with the Centre for Pattern Analysis and Machine Intelligence, Department of Electrical and Computer Engineering, University of Waterloo, ON, N2L 3G1, Canada (e-mail: milonislam@uwaterloo.ca*, and sheikh.nooruddin@uwaterloo.ca).

Fakhri Karray is with the Centre for Pattern Analysis and Machine Intelligence, Department of Electrical and Computer Engineering, University of Waterloo, ON, N2L 3G1, Canada and Mohamed bin Zayed University of Artificial Intelligence, Abu Dhabi, United Arab Emirates (e-mail: karray@uwaterloo.ca).

Ghulam Muhammad is with the Department of Computer Engineering, College of Computer and Information Sciences, King Saud University, Riyadh 11543, Saudi Arabia and Center of Smart Robotics Research, King Saud University, Riyadh, Saudi Arabia (e-mail: ghulam@ksu.edu.sa).

context-aware information processing employing network resources, are essential components of IoT [1], [2]. The IoT allows physical things to see, hear, think, and execute a variety of tasks by allowing them to "speak" to one another, exchange information data, and make choices. The concept of IoT turns these devices from being conventional to smart by leveraging its wide range of technologies including embedded applications, Internet protocol and applications, wireless communications, distributed networks, and sensor technologies [3], [4]. An important advancement of the current Internet into a network of interconnected objects is that it not only comes into contact with the physical surroundings via actuation, controlling, and monitoring, nor merely extracts information from the surroundings, but it also uses traditional network requirements to enhance data transfer, predictive analysis, and communication [4], [5]. The IoT covers a huge number of applications including healthcare, automotive, utilities, smart cities, wearables, smart homes, and smart farming [6], [7]. Numerous research projects, enterprises, and businesses are currently engaged in the creation of various IoT features to accommodate the growing evolving needs that emerge with such rapid development.

A healthcare platform is described as a combination of hardware and software components that work together to give a wide range of healthcare services and applications to people including healthcare professionals and patients, to promote health in an efficient and widespread manner [8], [9]. Nowadays, there are numerous healthcare frameworks available that integrate various technologies to monitor multiple human bio-physical symptoms and environmental data utilizing various wireless communications techniques including ZigBee, 5G, Bluetooth, and Wi-Fi [10], [11]. IoT technology supports healthcare systems by allowing people to remain at smart homes and be monitored in real-time rather than being admitted to hospitals or clinics that reduces the emergency costs [12], [13]. There are several advantages of IoT technologies to design smart healthcare systems such as ease of access, ensuring patient comfort and safety, and reducing patient's burden in hospitals. In general, IoT enables networks of smart devices, cloud applications, and solutions to make data transfer and storage easier. The most promising applications of IoT in the area of healthcare include remote monitoring, smart medical device, smart homes, as well as wearable devices. In recent years, several academic and industry researches on IoT interoperability including the healthcare domain have been done, with an emphasis on the standardization of communication protocols to offer interoperability of diverse devices, networks, and data structures [14], [15]. To ensure QoS in healthcare systems, IoT should incorporate some aspects such as standardization of reliable communication protocols, enhanced mobile and wearable devices, and low-cost as well as low-power embedded processors [16], [17]. The IoT communication architecture could be the primary enabling mechanism for decentralized pervasive healthcare applications, considering the prevailing availability of distant wireless healthcare platforms and the rising topic of electronic patientcare datasets [18], [19].

This paper is aimed at highlighting the fundamental issues of IoT device capabilities, architectures, protocols, and healthcare applications in detail. The IoT device incorporates all the necessary components for hardware and software platforms. The hardware platform describes the most popular portable processing devices such as Raspberry Pi, Arduino, and ESP8266 highlighting some significant features including power requirement, memory capacity, and processing power. The software platform is described focusing the operating systems for small embedded devices and built-in software for IoT-related tasks. The operating systems named Contiki, Mbed, RIOT, Embedded Linux, and Windows 10 IoT are described here with their distinguished features. The popular and widely used built-in tools for IoT smart applications include DeviceHive, Kaa, ThingsBoard, AWS IoT, and Google Cloud IoT. The IoT architectures such as three-layer, SoA-based, and middleware-based architectures are depicted to provide cutting-edge insights. The three-layer architecture contains the layers of perception, network, and application. Additionally, the SoA-based architecture comprises of a service layer along with the layers of the three-layer architecture. The middleware and business layers are the additional layer in the middleware-based architecture over the three-layer architecture. The IoT communication protocols are also demonstrated here to provide an overview of communication interfaces. The popular eight protocols that are described in this paper are CoAP, MQTT, XMPP, AMQP, DDS, LoWPAN, BLE, and Zigbee. The most promising healthcare applications for monitoring, diagnosis, and treatment for individuals using IoT technologies are demonstrated. The major applications include bio-physical parameter monitoring, chronic disease detection, medication management, telehealth, as well as home and elderly care. The current challenges for IoT technologies and potential future trends are explored to find out the research gaps and potential solutions to develop smart applications. The key contributions of the paper are highlighted as follows.

(i) Providing a comprehensive discussion on IoT device capabilities highlighting the area of hardware and software systems including the processing capabilities, operating systems, and built-in tools.

(ii) Addressing core technologies for IoT architectures that are the fundamental issues in diverse application domains of IoT and a detailed discussion on it.

(iii) Analyzing typical IoT communication protocols that deal with data transmission concerns via networks, as well as providing an overview of each.

(iv) Exploring the in-depth analysis of the most recent and promising IoT-based healthcare applications incorporating different architectures and protocols for monitoring, diagnosis, and treatment of the patients in hospitals as well as smart home environments.

(v) Presenting a study of the major open issues for application approaches in the IoT context, as well as the future trends to provide the scope of further research for smart applications.

The remaining parts of the paper read as follows. The paper collection and selection criteria are demonstrated in section II. Section III provides the concept of IoT device capabilities with proper illustration of hardware and software platforms. The architectures that are available in recent times to develop smart systems are described in section IV. Section V demonstrates the IoT protocols that are generally used to transfer data over the Internet. The healthcare applications of IoT technology are briefly described in section VI. Section VII demonstrated the open discussions of the IoT technology with the current challenges as well as the potential future works. Section VIII concludes the paper.

## II. PAPER COLLECTION AND SELECTION

This research was limited to peer-reviewed research works published in reputed international venues mostly within the years from 2017 to 2022 and written in English. Some notable sources used for this research are Google Scholar, EMBASE, PubMed, and NCBI. The most used search keywords were "IoT Overview", "IoT Security Issues", "IoT architectures", "IoT in Healthcare", "IoT Operating Systems", "Challenges in IoT", "Industrial IoT", and "Smart Applications using IoT".

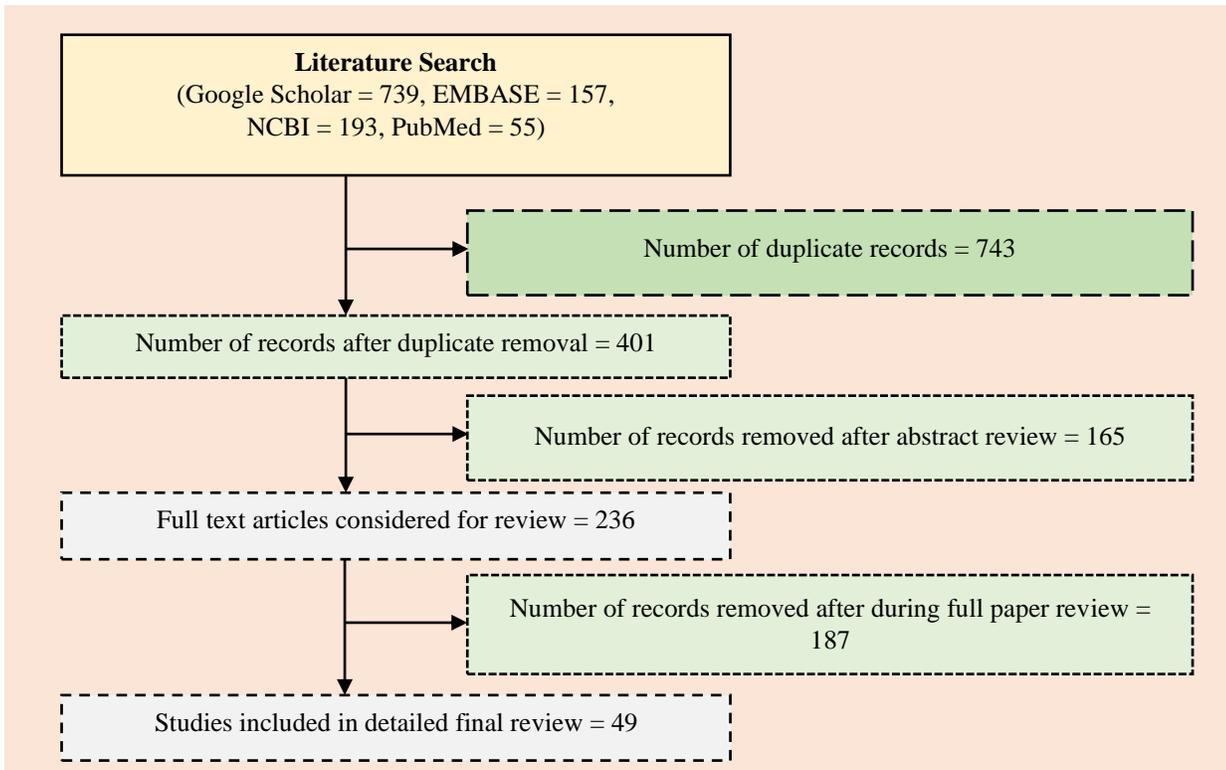

Fig. 1. Flow diagram showing article selection for final review. The final selection is done based on the duplicate records, abstract review, and full paper review.

We also manually searched various web search engines such as Google, DuckDuckGo, and Bing and books, articles, online magazines, online reports from reputable sources to provide various statistics. The abstracts, titles, keywords, and overall structures of the papers were analyzed to eliminate duplicates. We also eliminated publications that were beyond the scope of this work. Fig. 1 shows the article selection process of the detailed review paper from the initial search stage. Only a very small number of papers were selected for review from the initial high number of search results. The rejected papers were removed based on duplication, and out of focus of this work.

## III. IoT Device Capabilities

In general, an IoT framework comprises of devices that perform sensing, actuation, control, and surveillance tasks. IoT devices are able to transmit data to other devices and applications over the Internet, or they can retrieve data from other connected devices and process it locally or on a cloud server after sending it to centralized servers. The tasks are completed on the local server or within the IoT architecture depending on some temporal and spatial restrictions such as memory, communication delays, deadlines, computational power, and speed of operation. An IoT device may have multiple interfaces for communicating with other devices, in both wired and wireless mode. The common interfaces that are equipped with most of the IoT devices are I/O interfaces for sensors, Internet connectivity interfaces, memory and storage interfaces, and audio/video interfaces. There are various types of IoT objects such as wearable sensors, accelerometers, cameras, smart watches, smart locks, Global Positioning

System (GPS), and LED lights that are available for smart applications. Mostly all IoT devices produce large amounts of data in various forms, which while analyzed by data analytics engines, carries important information that can be used to guide local or remote activities. The IoT device capabilities highlighting hardware and software platform is shown in Fig. 2. In the following section, the device capabilities including hardware and software platforms are demonstrated in brief.

### A. Hardware

The hardware components for the Internet of Things have arrived in a variety of shapes and sizes, depending on the tasks at hand. For instance, the core CPUs are used to operate the phones, the sensors are utilized to perceive data from the physical environment, and the data are processed and analyzed on the edge devices. In the case of wired mode, the hardware aspects are the essential components of the IoT platform, and the capabilities of these devices have only grown in significance as IoT has progressed. The main hardware kits that are widely used to develop IoT applications are described here focusing on some properties such as processing power, memory capacity, and power requirement.

### (i) Raspberry Pi

Raspberry Pi is a series of single-board computers that compress a lot of processing power into a tiny box and offer desktop-like functionality [20]. The components of Raspberry Pi vary based on its version. In Raspberry Pi B+ model, there are four USB ports, two ports for the camera, an Ethernet port, an audio jack port, and a HDMI port. To implement real case studies, there are 17 General Purpose Input Output (GPIO) pins that are available to connect with any electronic components such as LEDs, sensors, switches, and motors. The minimum required logic level for the Raspberry Pi is almost 3.3V. In general, it is run over the Raspbian operating system. Additionally, it is also compatible with other operating systems such as Lakka, RaspBSD, and RetroPie [21]. The programming languages that are supported in Raspberry Pi are Python, Java, C++, and C. The combination of GPIO pins with the Linux operating system gives this device a lot of flexibility and gives users a lot of choices. The CPU speed of the standard Raspberry Pi model is 700 MHz to 1.4 GHz, memory capacity is 256 MB to 4 GB, and the power requirement is 2.5 Amp. These measures are varied based on the version of the Raspberry Pi model. The Raspberry Pi is well-suited for a wide range of applications due to its high connectivity, ease of use, portability, low power consumption, high flexibility, and high transferability [22]. However, there is no power button to keep it in sleep mode while it becomes idle, no built-in ADC, and replicability is quite hard for this device.

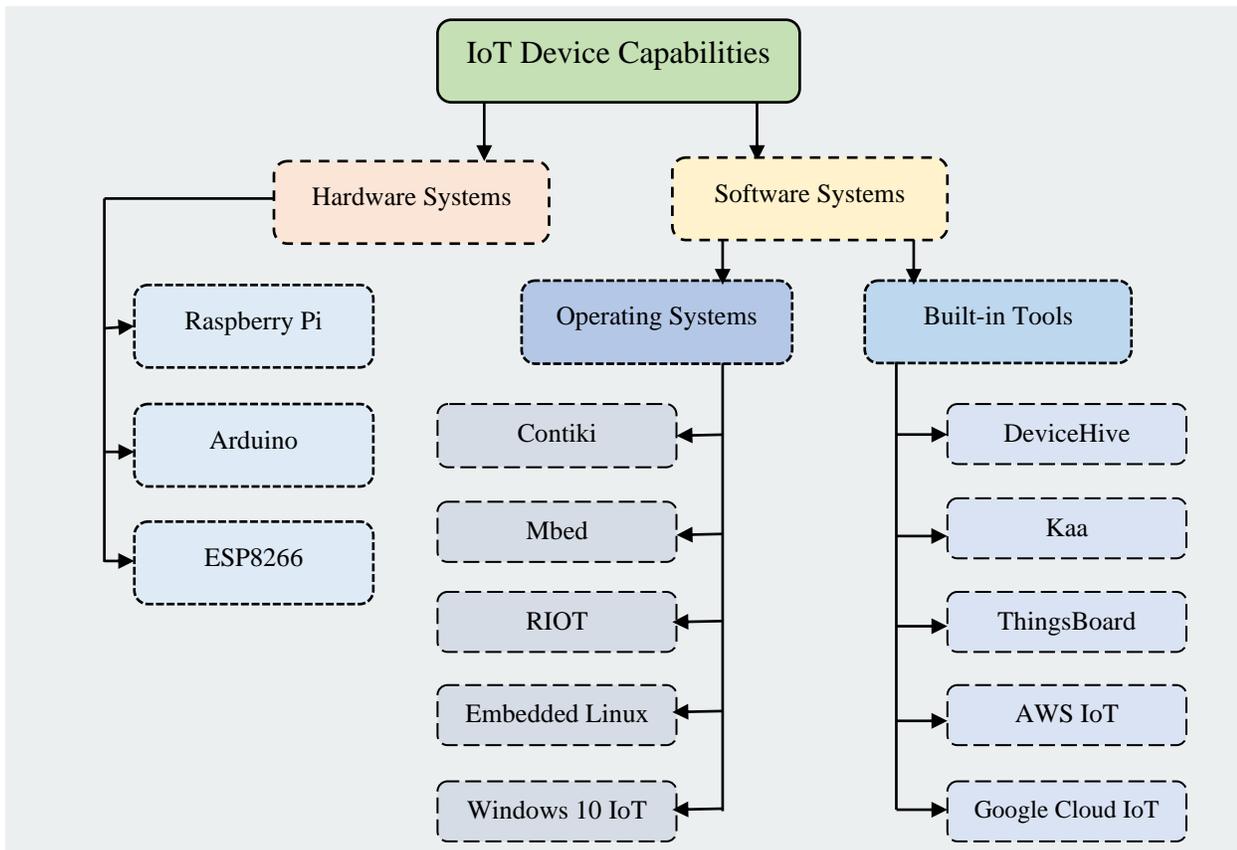

Fig. 2. IoT device capabilities considering hardware and software platforms. The hardware platform includes several types of tiny processing devices such as Raspberry Pi, Arduino, and ESP8266. The software module comprises several operating systems and built-in tools.

### (ii) Arduino

Arduino is a popular and frequently used open-source electronics framework based on simple hardware and software that can be utilized to develop electronic projects. It offers a wide variety of products, from straightforward 8-bit microcontroller boards to wearable technologies, IoT components, and 3-D printing [23]. The set of instructions to complete any tasks are executed through microcontrollers.

There are several versions of Arduino such as UNO, micro, pro, mega, and nano available in the market for electronic system development where the use of the Arduino model varies based on the nature of the applications. It is easily compatible with MAC, windows as well as Linux platforms. A type B USB cable is utilized to do programming in the Arduino board with the support of an IDE [24]. The supported language for Arduino is C++ where the fundamental issue of peripherals such as sensors and actuators are controlled through the use of header files. Arduino Uno has analog pins, digital pins, a USB connection, a reset button, a crystal oscillator, and a power jack. Arduino is highly suited to perceive data from physical environments i.e., analog data [25].

### (iii) ESP8266

ESP8266 is a mini-sized, Wi-Fi enabled device with full TCP/IP stack and microcontroller capability which is frequently used for embedded development in IoT environments [26]. This device can host an application that offloads all Wi-Fi network services to another underlying hardware. The Wi-Fi module of ESP8266 is pre-programmed with an ESP-AT command set firmware that enables any Arduino device to get a Wi-Fi connection. It runs at 80 MHz and uses a 32-bit RISC CPU (reduced instruction set computer) based on the Tensilica Xtensa L106. The boot ROM, instruction RAM, and data RAM size of ESP8266 are 64 KB, 64 KB, and 96 KB respectively.

| Property | Raspberry Pi | Arduino | ESP8266 |
|---|---|---|---|
| Programming | Python/Java/C++ | C++ | C++/Lua |
| Wi-Fi | USB dongle | Shield or ESP8266 | Built-in |
| Code Distribution | In-Situ | USB/SPI/Serial | Serial/OTA |
| Storage | SD-card | Built-in | Built-in |
| I/O | 17 GPIO | Arduino UNO: 13 GPIO/6 ADC | 10 GPIO/1 ADC |
| CPU speed | 700 MHz to 1.4 GHz | 16 MHz | 80 MHz |
| CPU | ARM Cortex-A7, ARM Cortex A-53, ARM Cortex-A72 | Atmel AVR, ARM Cortex-M0+, ARM Cortex-M3, Intel Quark | Tensilica L106 32-bit RISC processor |
| Memory Capacity (RAM) | 256 MB-4 GB | 16 KB-64 MB | 128 KB-4 MB |
| Power Requirement | 5V, 2.5 Amp | 5V, 250 mA | 3.3V, 70.5 mA |
| Cost | Expensive ($35) | Moderate ($3-$30) | Low cost ($3) |

The Serial Peripheral Interface (SPI) accesses the external flash memory. ESP8266 is quite fit for the low-cost systems [27] that do not require a high amount of CPU support, huge storage, some special features such as camera, USB, and HDMI ports.

Table I shows the different properties including CPU speed, memory capacity, power requirement, supported programming languages, and storage of the above-mentioned hardware components.

### B. Software

In general, IoT software outlines the key aspects of communication networks and action via numerous channels, integrated devices, partner systems, and development tools. IoT software is the underlying platform that allows a connected device to capture and communicate data in real-time where the data are generally processed by computers and relevant applications. The connected devices can operate effectively owing to the intelligence that IoT software ensures. Smart devices are limited to their computing power and in such scenarios, intelligence is ensured through cloud computing technology where the IoT software is located. The operating systems along with built-in tools which are the parts of IoT software are described here.

#### (i) Operating Systems

An IoT Operating System (OS) is a part of software platforms that ensures how IoT smart devices interact with users and handle all software and hardware resources. IoT OSs are integrated operating systems that are intended to work in small IoT devices with limited memory and low processing capacity. IoT devices can interact with cloud platforms and other intelligent devices over a global network thanks to embedded operating systems. The operating systems for a particular application are chosen based on some performance measures including flexibility, compatibility, reliability, simplicity, and consistency. The popular IoT operating systems are demonstrated here to provide a short overview of embedded operating systems.

#### 1) Contiki

Contiki is a freely available operating system that is well-known for its way of interacting with extremely small, low-power, and cost-effective embedded systems to the Internet [28]. The programming language that is used in Contiki is C.

It supports multithreading via protothread, and the processes are managed through cooperative or preemptive scheduling. Generally, multiple stacks are supported in Contiki including some features such as 6LoWPAN, CoAP, and IPv6 [29]. This operating system is appropriate for developing complex wireless systems due to its high memory efficiency.

#### 2) Mbed

Mbed operating system is a publicly available, open-source OS that is quite popular for its integration with ARM processors and has a wide range of connectivity features such as Wi-Fi, Ethernet, 6LoWPAN, Cellular, and Bluetooth [30]. It is particularly tailored for 32-bit ARM processors. This operating system has a multilayer security system that provides users with a high level of reliability. Mbed OS is well-suited to IoT applications on ARM cortex M-based devices, as it preserves code clean and portable features [31]. It is a popular IoT OS due to its low system requirements and compatibility with a variety of development kits.

#### 3) RIOT

RIOT is another widely used and popular operating system specifically designed for embedded devices [32]. This OS targets wireless sensor networks as it is developed based on a microkernel architecture named FireKernel. It ensures the usage of multithreading and SSL/TSL libraries, as well as provides the facilities of the use of 8, 16, and 32-bit processors. The supported language for RIOT is C and C++. RIOT can run as a Linux or macOS process as there is a dedicated port for this OS. RIOT enables modular design and provides access to separate hardware abstraction for minimal memory and energy consumption [33].

#### 4) Embedded Linux

Embedded Linux is a significantly modified version of the Linux kernel, designed specifically for embedded devices. The compact size and low power consumption of this OS make it well-suited for fulfilling all the requirements of IoT devices such as tablets, navigational devices, and wireless routers [34]. It requires a small amount of storage like 100 KB of memory that makes it quick and dynamic. In the IoT OS world, this OS also offers an unparallel level of customization. Embedded Linux OS is easily compatible with any single-board computer such as Raspberry Pi (Raspbian, Ubuntu Core, OSMC, and Gentoo. Linux distros are popular options for IoT applications

| Operating Systems | Features | Provider | License | Processor | Use Cases |
|---|---|---|---|---|---|
| Contiki | Open-source, free | ThingSquare | 3-clause BSD | ARM Cortex-M, MSP430, AVR, and x86. | Networked memory-constrained systems |
| Mbed | ARM-based, high-grade security | ARM | Apache 2.0 | ARM Cortex-M | For portable code |
| RIOT | Open-source, full multithreading | FU Berlin | LGPLv2.1 | ARM Cortex-M, MSP430, ARM7, AVR, x86, and Cortex-M23. | Can be run as a MacOS process |
| Embedded Linux | Linux kernel, free | Multiple providers | GPL, and GPLv2 | ARM Cortex-A8, ARM Cortex A-53, and Intel Core i3, i5, i7. | Versatile-can be used for various use cases |
| Windows 10 IoT | Proprietary, high-grade security | Microsoft Corporation | Commercial | ARM Cortex-A7, Snapdragon 400, ARM Cortex-A53 | Ideal for heavy-duty industrial use cases |

due to some performance factors such as flexibility, consistency, architectural layers, cloud support, and visualization [35].

### 5) Windows 10 IoT

Windows 10 IoT is a product of Microsoft Windows 10 operating system, but it is dedicatedly designed for IoT devices [36]. Windows IoT is divided into two parts: Windows 10 IoT core, which is considered to interact with tiny embedded devices, and Windows 10 IoT Enterprise, which is developed for industrial use. Windows 10 IoT includes a user-friendly interface, more user control than some other OSs, and is universally recognized as a capable IoT operating system among IoT experts. Windows OS is a quite better choice over free or open-source ones whenever reliability and security become critical issues [37].

Table II provides an overview of the discussed IoT operating systems including some dominant features such as accessibility, provider, license, processor, and use cases.

### (ii) Built-in-Tools

Generally, IoT software has two main attributes such as (i) Connectivity with wired or wireless devices via Bluetooth, ZigBee, Ethernet, 5G/6G, or satellite communications. (ii) Automated communication between the hardware, devices, or machinery without human involvement to initiate an operation or carry out actions. The available built-in tools are developed based on these two characteristics to carry out automated data analysis and transfer, as well as autonomous response. The most popular IoT built-in tools are described here to provide in-depth knowledge to the researchers.

### 1) DeviceHive

DeviceHive is a publicly accessible platform that offers communication and management tools for connected devices. There are several things included in DeviceHive such as communication layer, control software, multi-platform modules and users to launch smart home development, remote control and monitoring tool, and telemetry [38]. DeviceHive is compatible with Python, Node.js, and Java client modules. It uses the REST, WebSocket, and MQTT protocols to connect with other devices. This platform is a modular, hardware, and cloud independent micro service-oriented solution with access control APIs in several protocols that enable users to monitor, control, and analyze the behavior of their devices.

### 2) Kaa

Kaa is developed for enabling smart devices to communicate with each other across a big cloud. Kaa is an open-source development tool, and it is hardware-oriented i.e., it is well-suited for any hardware components such as sensors, gateways, and devices. Kaa's multi-purpose middleware enables users to develop IoT solutions, application support, and a variety of intelligent devices [39]. The set-up of Kaa tool is quite easy and it provides a large number of features that can be integrated into the platform easily. Kaa is a scalable IoT platform that uses a microservice design to provide obvious layers of abstraction, flexibility, and scalability [40].

### 3) ThingsBoard

ThingsBoard is a free server-side built-in tool for data collection, processing, monitoring, and device management [41]. It is deployable in any smart application including personal and commercial purposes. ThingsBoard supports several protocols such as CoAP, MQTT, and HTTP. The IoT devices are easily connected to the platform via ThingsBoard gateway. Although the back-end of ThingsBoard is written in Java, it supports several micro services based on Node.js. Some of the major reliant features of ThingsBoard are scalable, fault-tolerant, robust and efficient, adaptable, and resilient [42]. In practical scenarios, it is well-suited for predictive analytics tasks such as smart farming, smart metering, and smart energy.

### 4) AWS IoT

The Amazon Web Services (AWS) IoT tool is a reliable and secure development platform for industrial, business, and retail solutions. This platform can be used to analyze and manage data from Internet-connected devices and sensors in residences, businesses, healthcare, automobiles, and other locations [43]. It supports HTTP, LoRaWAN, the lightweight communication protocol, and MQTT to publish and subscribe to messages. This tool incorporates layers of security like encryption, secure access, constant monitoring, and audits. AWS IoT is able to handle smart objects and trillions of signals, and it can securely and efficiently interpret and transmit those signals to AWS endpoints and other devices [44].

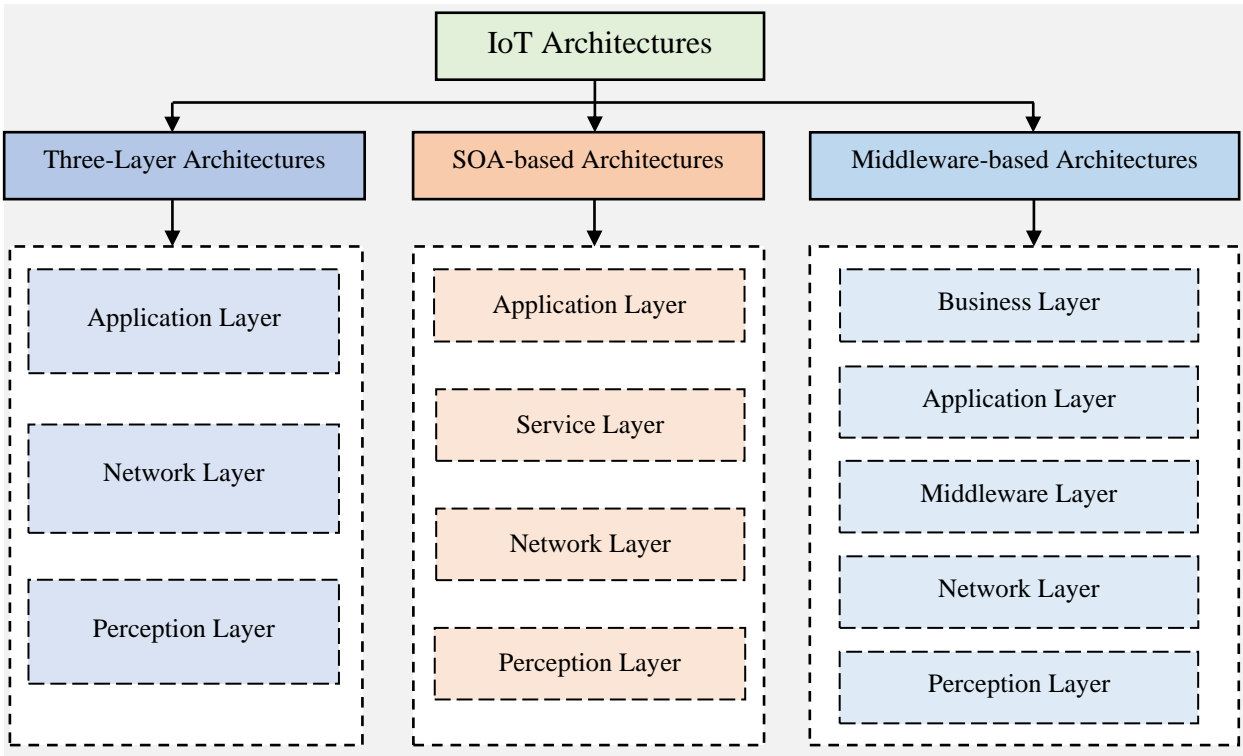

Fig. 3. Most common and frequently used IoT architectures that incorporate the three-layer architecture, SOA-based architecture, and middleware-based architecture.

### 5) Google Cloud IoT

Google Cloud provides scalable services that are quite managed and incorporated with IoT devices. The functionalities of this tool include connecting, preserving, and investigating data at the edge and in the cloud. The common aspects of Google Cloud IoT are data warehousing and quick querying, streaming and batch analysis, device connectivity and management, as well as Artificial Intelligence (AI) and Machine Learning (ML) models [45]. This platform is capable of predicting equipment maintenance and optimizing operations, as well as monitoring resources in real-time. Google Cloud IoT covers a wide range of applications including predictive analytics, logistics and supply chain management, smart cities and infrastructures, as well as real-time asset surveillance [46].

## IV. IoT ARCHITECTURES

The Internet of Things should be competent at networking billions or trillions of diverse things over the Internet, therefore, a flexible architectural pattern is crucial. Some of the major factors like reliability, scalability, privacy and security, quality of service (QoS), and data storage should be considered at the time of developing the architecture of an IoT network [47], [48]. To fulfill the demand of the defined key features, several research groups have proposed several IoT architectures. The ever-growing number of potential architectures has been unable to settle on a standard model [49]. There is no specific base architectural pattern till now despite numerous standardization initiatives. This is because of its wide area of applications having different design patterns and variables/factors to measure the performance of the existing systems. Fig. 3 depicts the most common and widely used IoT architectures, which are explained as follows.

### A. Three-Layer Architectures

The most fundamental architecture is the three-layer architecture which was developed in the beginning periods of research in the IoT area [50], [51]. This architecture is divided into three base architectures such as perception layer, network layer, and application layer which are demonstrated below.

### 1) Perception Layer

In the IoT ecosystem, the perception layer is sometimes referred to as the sensor layer, and it is deployed in the bottom layer. The perception layer interferes with the environments via smart devices like sensors, actuators, cameras, and GPS terminals to collect and process data. This layer's primary functions include connecting devices to the IoT framework, measuring, and processing real-time physical parameters (data) from connected things via intelligent devices, and transmitting data to the higher layer via its interfaces [52], [53]. In the perception layer, heterogeneous devices are configured using standardized plug-and-play techniques [54]. The major success of IoT, known as big data, begins in this layer.

### 2) Network Layer

The network layer, also known as the transmission layer and incorporated in the middle layer of the IoT framework, is referred to as the core of the IoT ecosystem [55]. The network layer serves as a link between the perception layer and the application layer, allowing data to be transferred to IoT servers, applications, and devices through interconnected networks. The network layer consists of different communication standards like Wi-Fi, Bluetooth, and ZigBee [56] and IoT objects like gateway, hub, and switching [57]. The use of wireless protocols is very important in this layer because wireless devices can be deployed in critical environments with less material and human effort [58]. This layer generally transmits/receives data to/from various applications via interfaces/gateways using a variety of communication standards and protocols.

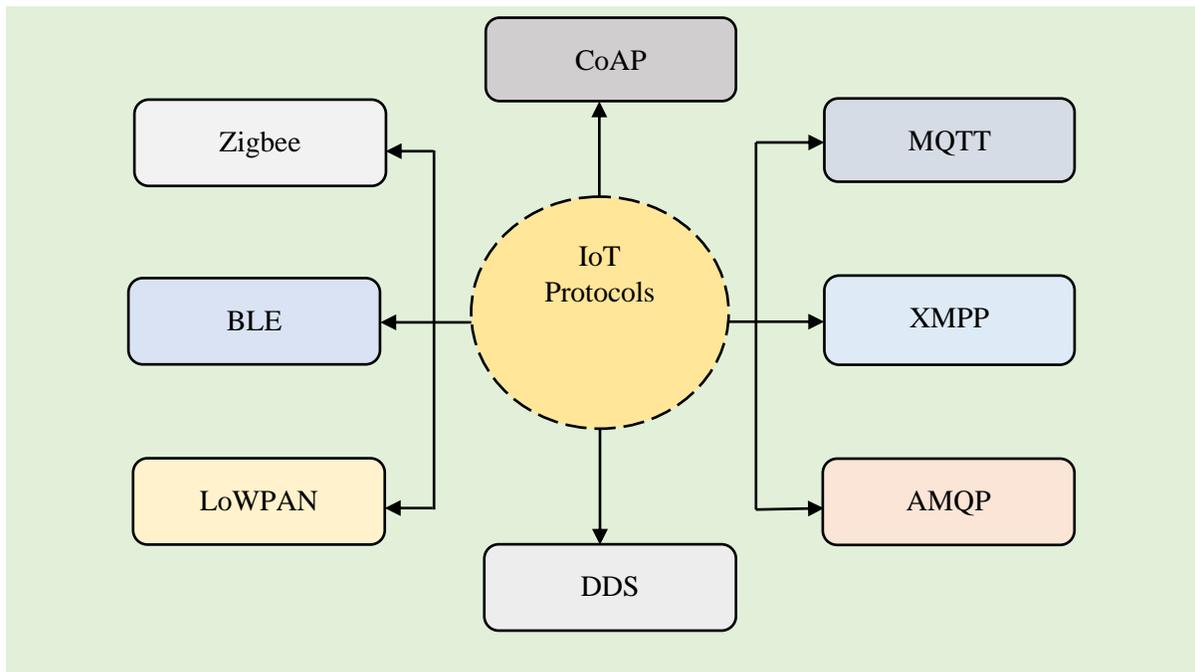

Fig. 4. Most common emerging IoT communication protocols that are extensively used to develop smart IoT applications.

### 3) Application Layer

The application layer is located at the top layer of the IoT network, and it serves as the interface between users and applications. This layer is in charge of retrieving data sent from the network layer and utilizing the collected data for specific operations [59]. As a task of the application layer, it provides the storage facility that can be utilized to store a backup of the acquired data into database. Additionally, this layer predicts the future condition of physical things by analyzing the nature of the received information. From an application perspective, this layer incorporates the functionalities of the IoT framework to develop real-time applications including the healthcare system [60], smart cities [61], smart transportation [62], and smart agriculture [63].

### B. Service-Oriented Architectures

Generally, Service-Oriented Architecture (SOA) is a component model-based framework that connects various functional blocks of applications via interfaces as well as standard protocols [64]. SOA provides viability by allowing software and hardware components to be reused and coordinating the intended services. A new layer named service layer is added between the network layer and application layer that ensures services are sufficient to assist the application layer extending the three-layer architecture in the SOA framework [65], [66]. As a result, the SOA architecture differs from the conventional three-layer framework by including service layer functionalities. The service layer consists of several components such as service management, service discovery, service composition, and service interfaces [67]. The service management maintains and measures the trust processes in order to fulfill the service's requests. The service discovery unit finds the service's requests. The purpose of service composition is to communicate with the associated things and incorporate services to obtain the requests efficiently. Additionally, all the services are interfaced with one another through service interface unit. The application layer, in general, complies with the user's request for services.

### C. Middleware-based Architectures

The middleware-based architecture is also popular and widely used in the IoT ecosystem nowadays; it is also known as a five-layer architecture. To ensure some of the performance factors like reliability, scalability, and QoS in today's applications, the middleware-based architecture plays a vital role by creating applications efficiently and providing links among the applications, users, and data [68]. The five-layer architecture has been introduced due to the recent developments of IoT and its huge applications in business models. As shown in Figure 3, the five layers include the middleware layer in between the network and application layer as well as the business layer at the top of the layer of the three-layer architecture [69]. The middleware layer aggregates, and filters data retrieved from the hardware, conducts data exploration, and offers access control to the objects for a variety of applications [70]. This layer has grown in importance in recent years since it allows for the growth of streamlined novel services and the incorporation of existing methods into new designs. The other layer of middleware-based architecture named the business layer organizes the entire services and activities of the IoT infrastructure. This layer takes data from the application layer and creates a business model, graphs, and other representations based on it [71]. This layer is in charge of designing, implementing, and monitoring all IoT components. In addition, the output of the previous four layers is compared with the desired output in this layer to enable the services and user's privacy [72].

## V. IoT Protocols

Various standards have been introduced to assess the services and relevance that are utilized for IoT solutions to link several things to the Internet in IoT common standards [73], [74]. Although multiple protocols have been developed, they are not all required for a single IoT application at the same time. The IoT protocols for a given application are chosen considering the nature of the application. The most common IoT protocols (depicted in Fig. 4), which are utilized in a variety of applications, are listed below.

### A. Constrained Application Protocol

The Internet Engineer Task Force (IETF), Constrained RESTful Environments (CoRE) research team developed the Constrained Application Protocol (CoAP), which is a Hypertext Transfer Protocol (HTTP) functional and lightweight application layer protocol [75]. As most IoT devices have limited power and storage, the CoAP protocol extends the functionalities of HTTP (which has a relatively high complexity) by fulfilling the needs of IoT devices [76]. This protocol shows how to build a web transfer protocol called REpresentational State Transfer (REST) on the upper level of HTTP. The CoAP uses the User Datagram Protocol (UDP) because it is simple in nature and has a small message size and layout, which helps to decrease the needs of bandwidth, utilize resources, and decrease the overhead of Transmission Control Protocol (TCP) handshaking before data transfer [77]. This protocol has two sub-layers such as the messaging sub-layer and the request/response sub-layer. The first sub-layer (messaging) determines the replications and ensures efficient data transmission over the UDP through exponential backoff as UDP is constrained to error recovery technique. On the contrary, the REST communications are handled by the request/response sub-layer. There are four kinds of messages in CoAP such as confirmable, non-confirmable, reset, and acknowledgment. The CoAP enables efficient delivery, congestion control, and flow control for IoT applications in resource-constrained and unsynchronized objects [78]. The CoAP has several drawbacks, including increased communication delay, packet delivery instability, and the inability to transfer complicated data [79].

### B. Message Queue Telemetry Transport

Message Queue Telemetry Transport (MQTT) is a messaging transport protocol that performs data aggregation of the environmental data and sends it to a web server [80]. This protocol is based on the TCP subscribe and publish messaging model and is intended for lightweight Machine-to-Machine (M2M), server-to-server, and machine-to-server interactions [81]. Here, the clients are acted as publishers/subscribers and the server is as a broker where the clients are connected to the server through TCP. Generally, the subscriber registers for a particular task in a device, and the data are generated and transferred to subscribers by the publishers through brokers [82]. MQTT is appropriate for utilization in things with limited resources, like those with low power and computing capabilities and are connected to low bandwidth or unstable networks. However, the MQTT protocol is not suited for usage with all forms of IoT applications because it operates over TCP, and the overhead is raised because it uses topic names as texts [83].

### C. Extensible Messaging and Presence Protocol

Extensible messaging and presence protocol (XMPP) is a protocol that ensures low bandwidth communication and short message transfer, making it ideal for video conferencing, publish-subscribe systems, telepresence, multi-party chatting, and talking in IoT [84]. For instant messaging applications, XMPP is appropriate for authentication, security measure, access control, hop-by-hop and end-to-end encryption, and interoperability with various protocols. This protocol serves three functions: client, server, and gateway, and it facilitates two-way communications between any two of these roles [85]. In this scenario, the client connects to the server through TCP protocol and transfers data using the XML streaming standard; the server is in charge for the connection management and routing of the message, and the gateway ensures reliable connectivity among distributed systems. This protocol allows communication among a variety of applications as it is flexible and simple in nature. However, XMPP requires high computing capabilities devices, consumes bandwidth of the network, transfers simple types of data, and is unable to provide QoS [86].

### D. Advanced Message Queuing Protocol

Advanced Message Queuing Protocol (AMQP) is an open platform messaging standard which is utilized at the application level to provide message services such as privacy, queuing, durability, and routing [87]. AMQP ensures reliable and consistent information exchange by using message passing primitives such as one-to-one, one-to-many, and exactly-once delivery. This protocol necessarily needs a stable transport protocol architecture and middleware serves as a gateway between applications and available resources, connecting institutions and mechanisms throughout time and space. The message queue and the exchange queue are two main steps in the AMQP data transmission process. In a message queue paradigm, messages are kept until they are delivered to the recipient. The messages are transmitted in an appropriate sequence in another scenario (exchange queue model) [88]. AMQP also enables the publish/subscribe communication architecture in addition to point-to-point data transfer. There are two kinds of messages found in AMQP like bare messages provided by the sender and annotated messages available at the recipient. However, AMQP requires comparatively higher bandwidth, and it does not guarantee resource discovery [89].

| Protocol Name | Network Standard | Latest Version (Year) | Transport Protocol | Messaging Model | Architecture | Security and QoS | Application Levels |
|---|---|---|---|---|---|---|---|
| CoAP | IETF, Eclipse Foundation | RFC 8323 (2018) | UDP | Request/response | Tree | Both | Utility field |
| MQTT | OASIS, Eclipse Foundations | MQTT version 5.0 (2018) | TCP | Publish/subscribe and request/response | Tree | Both | IoT messaging |
| XMPP | (RFC 3920-RFC 3923) RFC 4622, RFC 4854, RFC 4979, RFC 6122 | XMPP v 1.0.1, XEP-0128 (2019) | TCP | Publish/subscribe and request/response | Client-server | Security | Remote management |
| AMQP | OASIS, ISO/IEC | AMQP v 2.5.0 (2019) | TCP | Publish/subscribe | P2P | Both | Enterprise integration |
| DDS | OMG | DDS v.1.4 (2015) | TCP/UDP | Publish/subscribe and request/response | Bus | QoS | Military |
| LoWPAN | IEEE 802.15.4 | 6Lo-BLEMesh (2019) | TCP | Publish/subscribe | Star, mesh | Security | Structural monitoring |
| BLE | 802.15.1 | 6Lo-BLEMesh (2019), MRT-BLE (2018) | TCP/UDP | Publish/subscribe and request/response | Star | Security | Wearable devices |
| Zigbee | IEEE 802.15.4 | Zigbee 3.0 (2018) | UDP | Publish/subscribe and request/response | Star, mesh, hybrid | Security | Consumer electronics |

### E. Data Distribution Service

Data Distribution Service (DDS) is a data distribution service protocol based on publish/subscribe model that allows highly dependable M2M, real-time, flexible, and accessible connections [90]. DDS retrieves the edge outliers and gives notifications as well as pushes them to the data analytics engine. The distributed relational information model, which uses a distributed server to keep the application's specific data, is generated/consumed by the publishers/subscribers. The powerful QoS architecture of DDS is worked based on a series of two principles that provide end-to-end QoS control [91]. DDS allows asynchronous data transfer via a data bus through Peer-to-Peer (P2P) and distribution communication, making it a good fit for IoT applications [92]. As this protocol supports multicasting and broker-less model, it is possible to obtain high reliability and QoS through the use of this protocol.

### F. Low-Power Wireless Personal Area Networks

Low-Power Wireless Personal Area Networks (LoWPAN) are made up of a variety of cost-effective devices that are linked via wireless communication. This protocol has a huge number of applications in IoT architectures due to its small packet sizes, low computing power, low data throughput, and low latency [93]. Additionally, the 6LoWPAN protocol was introduced by incorporating the most recent release of the Internet protocol (IPv6) and LoWPAN. 6LoWPAN makes it easier to maintain the administration process by allowing each constrained object to be accessed independently within the network. Additionally, it is in charge of segmenting and reorganizing IPv6 traffic, guaranteeing unitary routing, reducing protocol stack headers, and providing compliance with the higher levels [94]. This protocol eliminates overall packet overhead as it does not include the extra header information during routing. Besides, 6LoWPAN contains a mesh address header to enable packet routing in a mesh architecture, but it is unable to provide detailed information of routing to the link layer. 6LoWPAN has various advantages, including ad-hoc self-organization, robust connectivity and standard compatibility, and low power consumption.

### G. Bluetooth Low Energy

Bluetooth Low Energy (BLE), an extended version of Bluetooth offers a small radio with reduced power consumption to operate for a longer period for controlling and monitoring applications [95]. The protocol stack that is utilized in BLE is almost identical to the standard of conventional Bluetooth technology, but it has a larger coverage; approximately 100 meters with low latency [96]. The devices that employ the BLE standard are categorized into two groups: master and slave. The master devices are the ones that play the most important roles and link to slaves. Additionally, the slaves can access and subscribe several master devices. BLE enables devices to investigate as masters or slaves channels in star topology [97]. This technology turns off the radio while idle time and only turns it on to broadcast or receive minimal data packets, resulting in minimal energy consumption. There is a gateway required (another BLE device with network connectivity) for BLE devices while transferring data over the Internet.

### H. Zigbee

Zigbee is a communication standard that ensures reliable, low power, and cost-effective data transfer, but it covers a small range of communication [98]. Zigbee supports star, cluster-tree, and P2P network topologies. A controller, in general, is in charge of the structure and can be found at the middle of a star network, at the root of a tree or cluster architecture, or anywhere else in a P2P topology [99]. There are two stacks in Zigbee standard such as ZigBee and ZigBee Pro; these stacks allow mesh network architectures and to operate with a variety of applications, allowing for low storage and processing power implementations. ZigBee Pro adds new functionalities like a symmetric key in order to ensure security, effective many-to-one relaying for enhanced performance and probabilistic address allocation for flexibility. This protocol offers stringent security and listening mechanisms, making it quite appropriate compared to other wireless communication standards [100]. The Zigbee protocol has a wide range of control and monitoring applications, including smart healthcare, smart homes, industry automation, lighting, and commercial control.

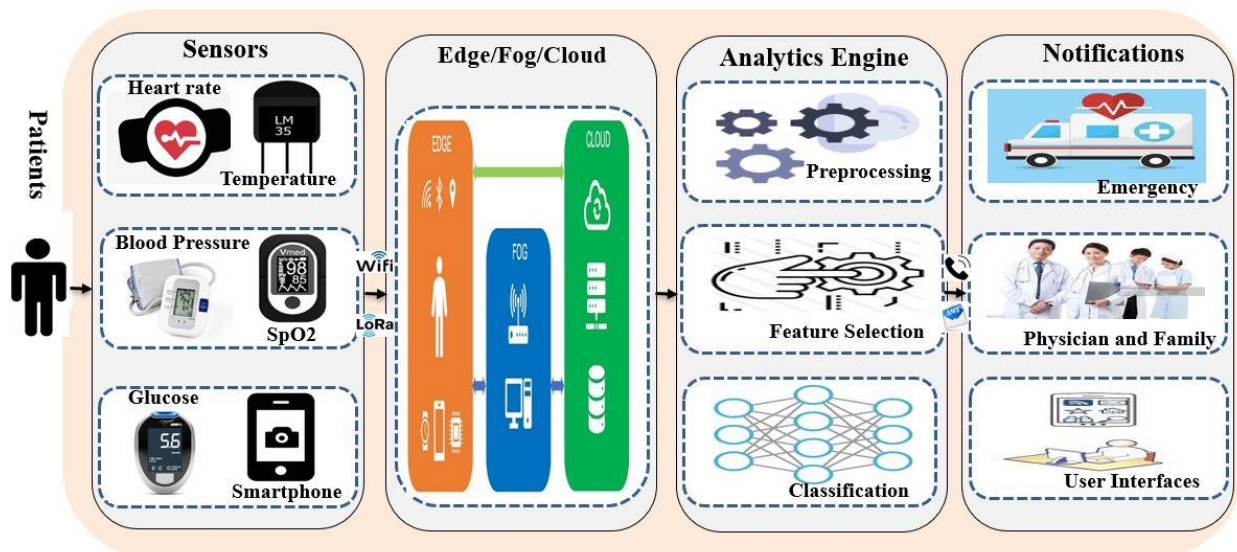

Fig. 5. A general pipeline of IoT-based smart healthcare systems. The sensors (heart rate, temperature, blood pressure, SpO2, glucose, and smartphone) collect the data from the real-time healthcare environment. Some of the data are processed in edge/fog server and the major data analysis occurs in cloud server where the data analytics engines predict the patient status. The emergencies, families, and physicians receive the patient's condition and take further required actions.

Table III describes the IoT communication protocols with some basic features such as network standard, latest version, used transport protocol, used messaging model, utilized architecture, the security and QoS measures, as well as potential application levels for the protocols that are necessary to compare each of the protocols.

## VI. SMART HEALTHCARE APPLICATIONS

Considering the state of the world today and the disperse of outbreaks and communicable diseases including COVID-19, as well as other things relevant to this global epidemic like massive cost, long distance, and the need for isolation during this crucial timeframe, heading to healthcare institutions is challenging, even sometimes become impossible particularly for the disabled and elderly, who experience severe from at least one chronic illness [101], [102]. As a result, efficient, substantial, and computer-aided technology is very important to address the necessities for long-term care and remote health care monitoring to ensure patients with an adequate quality of life while also reducing the economic strain [103], [104]. IoT has become a revolution for healthcare applications, allowing for the development of predictive and smart systems by linking various IoT devices to perceive real-time bio-physical information from patients including heart rate, blood pressure, glucose level, body temperature, and oxygen saturation level [105], [106]. The most significant contributions of IoT in the healthcare industry are remote patient monitoring, disease management, preventative care, and assisted living [107], [108]. Additionally, the application environment of IoT for smart healthcare includes hospital management, home healthcare, and e-healthcare/mobile healthcare [109], [110]. A general pipeline of IoT-based smart healthcare systems with the basic components is shown in Fig. 5. Various sensors/medical devices are utilized to collect data from the patient end through the help of healthcare professionals or the patients. The data collection step for healthcare is done in the perception layer and the retrieved data are sent through the network layer using Wi-Fi or LoRa communication technologies. Some portions of the collected data are processed in the edge/fog server to decrease the burden of the cloud server. The main data analytics including preprocessing, feature selection, and classification are done through predictive analytics engines with the use of artificial intelligence techniques; resided in the cloud server. All the processes for the data analysis are performed in the middleware layer. The outcomes of the analytics engines are sent as notifications through call/short messages to the emergencies, physicians and families, as well as interactive user interfaces. The application layer and business layer performed the information consuming step. Although IoT has diverse applications in the healthcare domain, some of the most recent and promising works are described here. The major applications that IoT includes in healthcare domain are depicted Fig. 6.

### A. Biophysical Parameter Monitoring

The monitoring of biophysical parameters is mainly used to observe the changes or absence of changes in the health findings of a patient. The goal of health monitoring is to provide early notification of threatening deterioration by achieving an optimal compromise that included many medical, engineering, and financial design factors. IoT plays a significant role to monitor patients remotely and condensing the burden of patients in the hospital as well as decreasing the number of cases. The major bio-physical symptoms of a particular patient involve Electrocardiogram (ECG), blood pressure, temperature, glucose level, and oxygen saturation (SpO2). The frameworks that are developed to monitor the above-mentioned parameters of the patients in the IoT environment are described below.

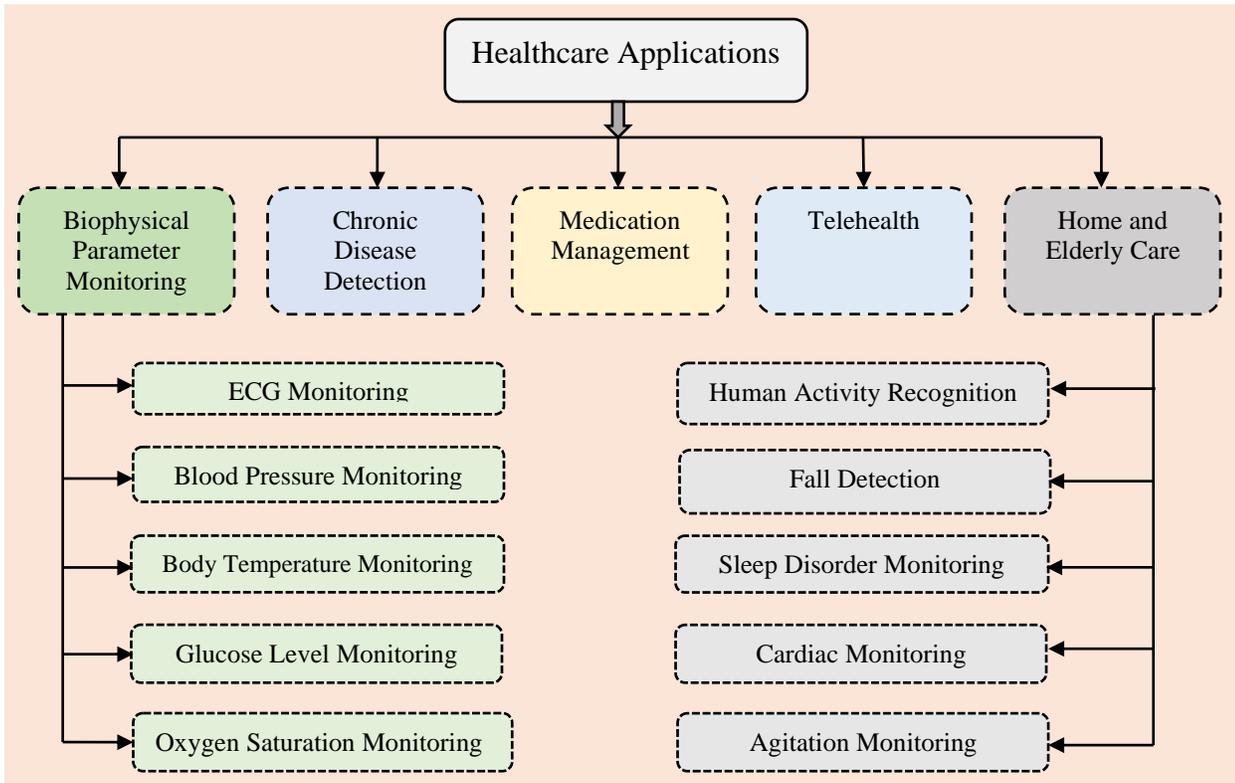

Fig. 6. Major healthcare applications focusing on IoT concepts in the area of biophysical parameter monitoring, chronic disease detection, medication management, telehealth, as well as home and elderly care.

### (i) ECG Monitoring

Wu et al. [113] introduced a tiny-sized and low-powered ECG monitoring framework that was incorporated into a t-shirt. To collect the ECG signal, the system used biopotential chips. The retrieved data from the patients are transferred to end-users (physicians/families) using Bluetooth. A smartphone application has been developed in this system to visualize the collected data. The minimum power required for the developed system to operate is 5.2 mW. The system is no longer in real-time mode; it is possible to monitor the ECG in real-time by utilizing big data analytics to accommodate larger data capacity. Shaown et al. [114] developed an ECG monitoring system in IoT for ensuring smart health. The developed system used Raspberry Pi, Arduino, ECG sensor (AD8232), and bio pad as hardware components. The collected data are sent to cloud server for processing through Wi-Fi network. The emergency cases of the individuals are automatically sent to doctors to take fast initiatives as a form of an email. The overall accuracy found from the developed system is approximately 80%. In another study, Djelouat et al. [115] presented a wearable ECG monitoring platform that resolved the problems related to power consumption. The device featured a novel design called compressive sensing, which can reduce power consumption and improve ECG monitoring efficiency. It is shown from the experimental results that the multicore processors are quite appropriate for a gateway-centric solution while maintaining processing capacity and power efficiency. Sahu et al. [116] proposed a cloud-based remote ECG monitoring scheme for cardiovascular patients. The ECG data are transmitted to the AWS through a mobile gateway service. BLE is used as a communication medium between the users and end devices. For data visualization, fast response, and long-live connection, AWS used HTTP and MQTT protocols in this system. Further, Basu et al. [117] introduced a platform for remote patients to monitor their cardiac conditions using the concept of IoT technology. The system used an ECG sensor (AD8232); interfaced with a notch filter (removes the power-line frequencies), and Arduino to process the data initially. The final processing is done using MATLAB tool to remove baseline wander and the processed data are transferred to the cloud as a decision via Wi-Fi that can be accessed by the registered users.

### (ii) Blood Pressure Monitoring

The measurement of blood pressure (BP) is one of the most important aspects of diagnosing any disorders in the human body. In general, trained professionals measure blood pressure while recording the data of humans. Nowadays, IoT and several sensing modalities are used to measure blood pressure automatically and efficiently. Numerous works [118], [119] have been conducted to make the blood pressure measurement process automated in recent years.

The authors of [120] developed a wearable device to measure systolic and diastolic pressure accurately. The developed system used light-based sensors to collect data and the recorded data are stored in the cloud server to make further analysis. The system was tested with targeted users (more than 70 individuals) and the performance was validated. Singh et al. [121] demonstrated a blood pressure monitoring framework using the concept of IoT, Arduino, and blood pressure sensors. Communication technologies including Bluetooth and Wi-Fi are utilized to transmit data from the patients to the cloud database. The developed system is flexible, cost-effective, compact, and provides the facilities to monitor blood pressure from the home environment. In another study, Hashim et al. [122] introduced a wearable system for blood pressure monitoring in IoT environment. A blood pressure sensor is connected with Raspberry Pi to collect data from the patients. The system is tested with various positions of the patients such as sitting, standing, and lying where the BP value of the sitting position is more correct. The readings of the BP sensor are sent to registered Telegram as well as Gmail application for patient monitoring. In [123], the authors presented the use of machine learning techniques to classify blood pressure in wireless sensor networks environment that can provide BP status and the abnormalities in real-time. In this prototype, a pressure sensor (MPS20N0040D-D) is used for data collection, a microcontroller (ATmega328) is utilized for data processing, and the data transfer is done through a transceiver (NRF24L01). Among the used machine learning techniques, the decision tree algorithm achieved the highest accuracy (97.9%) for BP status detection. Finally, all the data of BP status are retained in IBM Watson Cloud. Furthermore, Adi and Kitagawa [124] developed an IoT platform for blood pressure monitoring using a tiny embedded device. The used components in this prototype are the pressure sensor, Zigbee Radio Frequency (RF) module, and Raspberry Pi 3. The purpose of the Zigbee module is to send BP data to a database, which is then transferred for an IoT network via Zigbee end-device connection to the Zigbee coordinator. However, the developed system is still working in localhost; not in real-time.

### (iii) Body Temperature Monitoring

The body temperature of individuals is a leading determinant of homeostasis maintenance and is used in a variety of diagnostic procedures. In many disorders, maintaining records of temperature changes might help doctors to draw assumptions about a patient's health. The traditional way of temperature measurement is the use of thermometer that takes the records from the different positions of the patient's body such as ear, mouth, or rectum. However, the traditional techniques for body temperature measurement are low comfortable, less flexible, and accurate. In recent days, several frameworks [106], [125] have been developed thanks to IoT technology for human body temperature measurement.

Ota et al. [126] developed a wearable prototype for body temperature measurement that is printed using 3D printer and is applicable to wear on the ear. The developed device records data from the tympanic membrane utilizing an infrared sensor. In this system, a wireless sensor network and data processing unit are utilized where the prototype is integrated. The environmental conditions and other physical activities could not interfere with the readings of the body temperature. Zakaria et al. [127] proposed a lightweight device for real-time temperature measurement targeting infant patients. The major components of this system include a wearable sensor (LM-35), microcontroller (Arduino), ESP8266, and power module. The collected body temperature is visualized in the cloud server through ThingSpeak and Blynk application. The developed system can provide alert messages to the mediators to take immediate actions while there is a high change of temperature. In another work, Rahimoon et al. [128] introduced a contactless body temperature measurement system to monitor the patients remotely. Two temperature sensors named LM-35 and MLX-90614, Arduino, ESP Wi-Fi shield are used in this prototype. The collected data are transferred to the cloud server for remote access utilizing Wi-Fi connection. However, the developed prototype is quite accurate, real-time, as well as reliable. Boonsong et al. [129] demonstrated a contactless body temperature monitoring platform in IoT environment. The infrared temperature sensor MLX90614 DCI is used here to perceive the data from the environment and sent it to the host computer as well as cloud server using 2.4 GHz microwave frequency. The experimental findings revealed that the developed framework obtained a reliability calibration of 74.7%. Further, Fang et al. [130] used a wearable smart bracelet to monitor the temperature for contact tracing application during the time of the pandemic. The bracelet used RF technology for data collection and the retrieved data are sent to cloud database via wireless technology. However, the system is cost-effective, easy to operate, and provides better data privacy.

### (iv) Glucose Level Monitoring

Glucose level monitoring is one of the most significant aspects of managing type 1 or type 2 diabetes. Diabetes is a chronic disease wherein the glucose levels in the body keep going up for an extended time. Fingerpicking is the most common method for detecting diabetes, which is validated by blood glucose levels. The development of numerous wearable devices [131], [132] for blood glucose monitoring using modern IoT technologies ensures some features such as comfort, safety, noninvasiveness, and convenience.

Alarcón-Paredes et al. [133] developed a prototype in the shape of a glove that can measure a patient's glucose level and is equipped with a Raspberry Pi camera and visible laser beam. The developed framework used a set of images collected from the patients' fingertips to detect their diabetic status. For data analysis, an artificial neural network is used, and the obtained mean absolute error and Clarke grid error of this system are 10.37% and 90.32% respectively. However, the developed prototype is a little bit bulky to use. Gia et al. [134] introduced an architecture for real-time blood glucose level monitoring in IoT environment. The proposed scheme was designed as an energy-efficient sensor module using nRF communication protocol and developed an energy harvesting unit to extend the battery life of the sensor device. There is a push-button service in the developed prototype that notifies the mediators in the case of abnormal behaviors of glucose like too high or too low. In another research, Rahmat et al.

[135] proposed a prototype named GluQo to monitor the glucose level efficiently. The data from the fingertip are collected using near-infrared LED to determine blood glucose optically and the intensity of the retrieved light calculated the concentration of the glucose of the blood. The glucose level was sent to the mobile application through Wi-Fi signal from the microcontroller. It is found from the experimental findings that the percentage error of the developed prototype is 7.20%. Valenzuela et al. [136] used a double moving average technique enabled by an IoT network to offer a framework for blood glucose monitoring to minimize potential problems in elderly adults. The data for the developed system are collected using a glucometer and this sensor is connected with NodeMCU to make IoT environments. The system can notify doctors and families about a potential critical state in patients at an early stage. Further, the authors of [137] used near-infrared sensors to demonstrate a portable non-invasive glucose monitoring system. The data for the experiments are collected from the fingertip of the patients and the analyzed data are displayed on the Android application. The obtained percentage difference of the reliability test of the developed prototype is less than 25% and the approximate cost is $15.

### (v) Oxygen Saturation Monitoring

The measurement of oxygen saturation is a crucial thing in the management and interpretation of the healthcare system. Pulse oximetry is the process of measuring oxygen saturation in a non-invasive way and it is used as a significant factor in the patient care system. The non-invasive techniques for oxygen saturation monitoring solve the problems associated with traditional approaches and ensure real-time monitoring. The integration of IoT-based technologies [138], [139] with pulse oximetry has shown significant potential in the medical field.

Son et al. [140] proposed a SpO2 monitoring wearable tool that uses IoT technology to give real-time monitoring and prediction of patient health state. The data from the sensor are transferred to cloud database through Internet and this data can be accessed via a web server for the registered users. The experimental findings found that the highest deviation of SpO2 measurement for this device is approximately 4.4%. It is evident that the prototype is complete, cost-effective, and low-power consumption. The authors of [141] demonstrated a wearable prototype for blood oxygen saturation monitoring from Photoplethysmography (PPG) signal using the concept of Bluetooth low energy. Here, the system introduced a novel adaptive cancellation algorithm using adaptive filtering to remove motion-induced nosiness. The proposed architecture captured the blood oxygen signals as soon as it received the instruction and sent the data to the mobile phone gateway via Wi-Fi, where it displayed the mathematical waveform. In another research, Zhang et al. [142] proposed a new wearable tool for blood oxygen saturation monitoring that sensed two-channel PPG signals, and sent the signals to the smartphone wirelessly. The dataset for this work is retrieved from the users (eleven individuals) to measure the performance of the proposed framework. It is evident from the experiments that the root mean square error for SpO2 estimation is 1.8%. However, the performance of the proposed framework is calculated using the data from healthy patients. Adiputra et al.

[143] presented a small, wearable, and cost-effective SpO2 device in IoT environment for blood oxygen monitoring. To create IoT environment, the developed prototype is connected with NodeMCU which functions as a data processor and Internet network gateway. A dedicated website is developed to ensure online and real-time data accessibility from the cloud. The error rate for SpO2 monitoring is quite low (1.5% only). Furthermore, Utomo et al. [144] demonstrated a scheme for SpO2 monitoring using oximeter sensor data through the concept of IoT technology. The data are collected from the finger of the patients in different body positions like standing, sitting, and lying. The sensor is connected with ESP8266 microcontroller, and the collected data are transferred via the built-in Wi-Fi module to the cloud server where the data visualization is done using ThinkSpeak application. The experimental findings revealed that the uncertainly measure of the SpO2 is 0.04.

### B. Chronic Disease Detection

Chronic disease is a patient health issue whose consequences are continuous or otherwise lengthy, or an illness that develops over time. Many individuals around the world suffer from chronic illnesses including heart disease, diabetes, asthma, and cancer. In most cases, these illnesses lead to mental health problems, which is one of the most common adverse events of chronic disease. The recent revolution of IoT contributes a lot to develop healthcare frameworks [145], [146] for the people infected with chronic disease.

Shihab et al. [147] developed a system for heart disease detection using the concept of recurrent neural network (RNN) in IoT environment. In this system, an ECG sensor, Arduino, and biomedical patch are used for data collection and to perform the experimental study. The parameters from ECG recordings are retrieved and fed into RNNs with recurrent layers as input. The major advantage of this developed system is that it is quite beneficial for doctors to be able to prescribe medication to patients without having to be physically there. The authors of [148] proposed a IoT-based novel heart disease diagnosis system using machine learning techniques. In the preprocessing step, the noises from the raw ECG signals are removed efficiently ensuring more accurate diagnosis. As hardware components for ensuring IoT platform, the system is comprised of Arduino, ECG sensor, pulse sensor, and temperature sensor. The decision tree algorithm obtained comparatively better performances for heart disease diagnosis using the benchmark dataset. Hebbale et al. [149] presented a non-invasive self-care framework that monitors the blood sugar of diabetes patients using machine learning techniques and IoT. In this system, predictive analytics engines reside on the cloud, and they are used for data analysis for diabetes detection. A smartphone application facilitated the physicians as well as patients to monitor the blood sugar and related risk easily. The decisions from the machine learning approaches are forwarded to the doctors for further analysis and the necessary recommendations from the doctors are sent to the parents through the developed mobile application. Support vector machine obtained the highest accuracy of 82.41% from the experiment. The authors of [150] described an IoT-based diabetes prediction scheme using machine learning approaches. The data for this system are collected using IoT diabetes senor. The dataset for the experiment is divided into training and testing set in the percentage of 75% (11,250 records) and 25% (3,750 records) respectively. It is found that the achieved accuracy and error rate of the developed system are 94.23% and 5.77% respectively.

Shah et al. [151] proposed a health parameter monitoring scheme for asthma patients using the concept of IoT and cloud technology. The system utilized a smart sensor to record the respiratory rate of the patients. A watermarking technique ensures data security and authenticity whenever transferred from patients' side to cloud as well as cloud to physicians. A dedicated cloud server is used here to keep the record of the patients and it provides access to mediators for diagnostic and monitoring purposes. The author of [152] demonstrated an architecture to predict asthma using decision tree algorithm in IoT environment. The major components of the developed system are Arduino, ESP8266, GP2Y1010AU0F Optical Dust sensor, LM-35 sensor, and 'NO' Gas sensor. The prototype can notify patients when they need to take pain relievers. It is observed from the experiments that the decision tree algorithm achieved an accuracy of 97.7% for asthma prediction. However, the developed prototype is not fully integrated and bulky. Rehman et al. [153] developed an IoT cancer monitoring framework based on multi-layered architecture that can diagnose and monitor cancer patients remotely. Here, three-layered architectures are utilized which are the patient layer, connectivity layer, and medical layer. The prototype is in the shape of a wristband and can monitor the basic health issues of cancer patients. The major advantage of this architecture includes it gives continuous monitoring of the patients that is helpful for the physicians to generate a complete history of the patients which in turn would be helpful for treatment in the later stages. In [154], the authors demonstrated various types of architectures and frameworks to develop IoT-based healthcare platforms targeting cancer patients. For communication purposes, a wireless sensor network is developed that can accommodate a high number of sensors and the data analysis is done on the cloud server. The security measures and operational challenges regarding the IoT healthcare system for cancer patients are also described here.

### C. Medication Management

Smart medication is one of the most significant concerns in the IoT-based smart healthcare application. It is very critical as it is intrinsically related to human health, and missing a dose of medication can have serious effects. Medication nonadherence is particularly common in older individuals as they age and eventually increases the chance of diseases including dementia, and cognitive decline. Hence, it is badly needed for them to use assistive technologies to keep the track of medication timing and dose. In recent times, several studies [155], [156] have been performed emphasizing on employing IoT technology to track patients' medication adherence.

Bharadwaj et al. [157] developed an IoT device in the shape of a medical box to remind the patients in the time of medication. In this prototype, there are three trays to keep the medicine for three different periods such as morning, noon, and evening. The data are stored on the cloud server and smartphone application retrieved the data from the cloud and used it to remind the patients for medication. Srinivas et al. [158] introduced a home-based medicine box named iMedBox to manage medicine for patients in IoT environment. The used sensors in this prototype notified the patients via a wireless connection and an Android application to take their medications on time and to keep in closer contact with their doctors. In another research, Latif et al. [159] presented a wearable IoT device called I-CARES to diagnose health issues and medication for elder patients. The system described an algorithm for data analysis and aided the medical professionals to make decisions for early-stage treatment of predicted diseases or illnesses. Additionally, the developed prototype can predict emergencies and call emergency personnel when necessary. Tsai et al. [160] demonstrated an intelligent pillbox for the older individual and nursing home needs using the concept of electronic technology and network functionality. Using embedded sensors, the developed device detects the signal when patients take their medications and displays the schedules and deadlines as text, pattern, or voice on the LCD screen. The hardware unit is made up of webduino for data transmission and can interact with a web server via the IoT architecture. The sensed data are transferred to the Arduino module, which is then sent to the developed pillbox, allowing the elderly to maintain their medications. The authors of [161] designed a smart medication box for ensuring medication monitoring in IoT platform. The prototype includes six distinct parts to keep track of six different pills, and it can send timely reminders to the caregivers/patients through the smartphone. A simple authentication process is developed to avoid overdosage and improper intake of medicines. In addition, the system is able to provide monitoring facilities of temperature and heartbeat through the use of biosensors. Vardhini et al. [162] proposed an assistive system to help elderly people with memory impairment for keeping track of their medication schedules and inform them to consider taking their medicines at the appropriate times. The main hardware components for the prototype include NodeMCU, LDR sensor, motor, switch, power supply (battery), LCD monitor, and buzzer. The developed system is low-cost (approximately $15), and easy to use. However, the system has not tested in any hospital environments.

### D. Telehealth

Telehealth is the use of technologies such as digital devices, smartphone apps, and webpages to instruct, detect, and even prescribe remedy to individuals suffering from health problems. It primarily allows individuals of all ages to be monitored remotely. The telehealth system monitors the significant clinical data and takes oversight of the individual's well-being. Numerous frameworks [163], [164] have been done to ensure telehealth services using IoT technology in recent times.

Rokonuzzaman et al. [165] developed a telehealth platform based on the IoT concept to create warnings automatically for providing individuals with desirable healthcare under constant health contexts. A smartphone application called 'village doctor app (VDA)' is deployed for real-time monitoring. This scheme provides the monitoring facilities of heart rate and temperature. The hardware components of the developed prototype are heart rate sensor, temperature sensor, Arduino, NodeMCU, battery, and LCD display. The physicians can access the data from the cloud server through VDA application. Ganesh et al. [166] introduced a framework named 'AutoImpilo' to improve the telemedicine and telehealth services in IoT networks. The designed approach ensured that all online medical checks were conducted in collaboration with consulting the doctors or specialists. This system is quite beneficial for individuals who are unable to go outside due to ages or quarantine issues. The system used a smart card in the authentication step to ensure privacy, but it is not linked to the cloud. In another work, Tsiouris et al. [167] integrated many online technologies and compatible components into a unified framework for digital coaching, encouragement, and empowerment of elderly with balance problems. This approach has been proven in a telerehabilitation system for individuals with balance issues that uses augmented reality to provide a surrogate physiotherapist as a superimposed hologram, along with easy-to-use wearable sensors. Hewa et al. [168] employed many decentralized applications to transmit cryptocurrency to patients' wallets effectively and reliably as an incentive for providing their health information utilizing edge computing and blockchain technology. To provide scalability with a large number of connected biomedical sensors to the cloud, the system included memory offloading functionality into the blockchain. The proposed platform is tested in a near real-world setting employing the Hyperledger Fabric blockchain infrastructure and Raspberry Pi modules to imitate smart sensor activities. Further, Hamil et al. [169] described a telehealth platform for bio-signal diagnosis and classification using e-health sensors data considering IoT environment. The Arduino and Raspberry Pi with sensors are used for data acquisition and processing. The used sensors are here ECG sensors, body temperature sensor, SpO2, position sensor, airflow sensor, and galvanic skin sensor. Various machine learning algorithms are tested with PhysioNet databases where the best accuracy of 99.56% is found from convolutional neural network-support vector machine for atrial fibrillation classification.

### E. Home and Elderly Care

Elders can continue living at home freely with the assistance of smart home care services, as IoT devices provide them with the security of being observed and the capability to instantly call for help in the case of an emergency. Modern IoT technologies, as well as relevant wearables devices, are frequently utilized in ambient assisted living environments to monitor older patients who are unable to move rapidly and need longer time to reach clinics for routine or urgent medical facilities [170], [171].

Abdelgawad et al. [172] proposed a system for health monitoring of older people in an ambient assisted living environment where the data are retrieved from several sensors and transferred it to the cloud server for further investigation. The prototype incorporates six different sensors, Raspberry Pi, Wi-Fi module, cloud server, sensor interface circuits, and an indoor positioning module enabled by Bluetooth low energy. A feedback signal is sent to the users based on the behavior of the data so that they could take necessary actions in the case of emergencies. The authors of [173] developed a wearable framework for health data monitoring (stress, blood pressure, and position of the users) for the elderly in smart home environments. The stress of the elders is detected using electrodermal activity, PPG, and skin temperature sensors. The blood pressure is estimated utilizing the data from the PPG sensor. The location of the individuals is detected using a voice-assisted indoor location module. The highest accuracy of 94% is obtained from the combination of multiple signals for stress detection. Finally, a prototype is developed to test the location detection system in a smart home environment. In another work, Debauchea et al. [174] demonstrated a novel architecture for elderly psychological signals monitoring utilizing the concept of IoT and fog computing. The framework includes a wireless sensor network consisting of various psychological and environmental signals, a local gateway for local and fast storage, and Lamda cloud server for data processing as well as storage. The main advantage of this architecture is that it can provide local gateway monitoring of new and current patient data. The latency and CPU usage of the fog architecture is 7 ms and 7.86% respectively. In [175], the authors presented a novel and scalable platform for smart home and elderly care based on IoT concept. The system employed various sensors data such as heartbeat, pressure, and gas sensor for patients as well as smart home monitoring through the use of Raspberry Pi and cloud server. The medications are suggested by the registered doctors based on the health conditions of the patients which are delivered through wireless sensor networks in the form of texts or email. The experimental findings revealed that the mobility absorption of the designed network is approximately 90%. However, a few numbers of signals are considered here for health and home monitoring. Furthermore, Srinivasan et al. [176] used the concept of IoT and machine learning to develop an elder care system. The prototype of this framework contains sensors connected to an ESP32 and a home module with an IoT camera attached to a Raspberry Pi. The biophysical parameters are retrieved through sensors and processed in the AWS cloud utilizing machine learning approaches. The warnings are directed to caregivers automatically in the case of abnormalities. The locations are tracked through a GPS module that are transferred to the mediators to let them the location of the individual patient. All the data for the experiments are visualized through the developed smartphone application.

## VII. OPEN CHALLENGES AND FUTURE TRENDS

Internet of Things and the technology and research state related to it are in their nascent state compared to other networking research topics. The state-of-the-art connection models and architectures are recently being developed. IoT in general and IoT in healthcare in specific still face many challenges. Understanding and exploring these challenges such as the security and privacy risks, lack of standardization, big data management, and costs are imperative to realize the full potential of Internet of Things in industrial, healthcare, and personal use cases among other promising domains. The major challenges and the potentials future directions for IoT research are illustrated in Fig. 7 in brief. The open issues, as well as future research trends related to IoT, are discussed as follows.

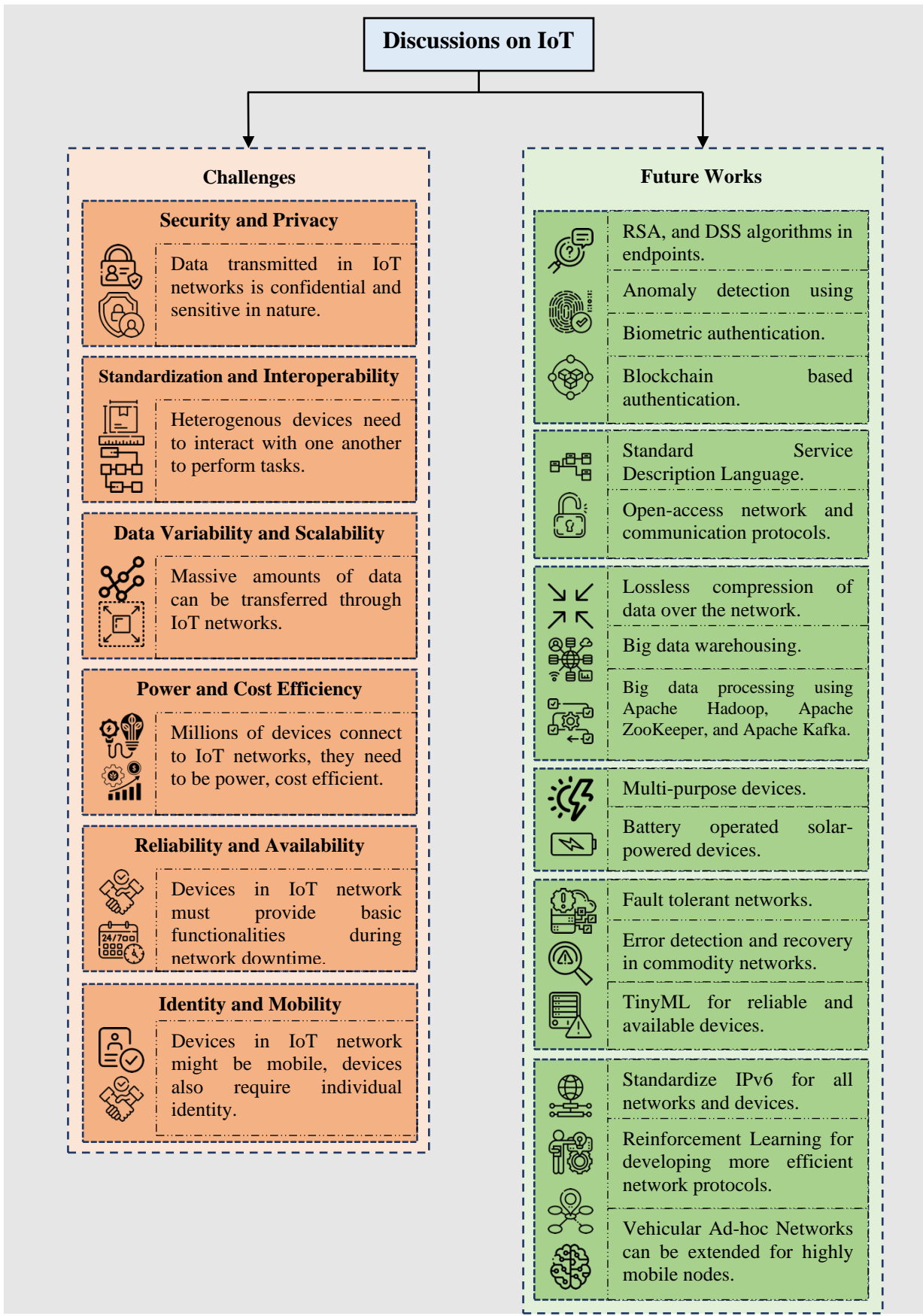

Fig. 7. Challenges and future research directions of IoT. The major challenges encompass security, reliability, scalability, availability, and cost-efficiency. The potential future directions include the area of blockchain, TinyML, reinforcement learning, and big data processing.

### A. Security and Privacy in IoT

Security is a huge concern in any networked system. IoT is a collection of devices connected via a single network or multiple networks. Thus, the security issues concerning networked systems are also relevant to IoT systems [177]–[179]. Devices in IoT are also heterogeneous in nature–which adds to the complexity in maintaining the security of the transmitted data. To provide complete security, all of the types of devices connected to IoT networks, would need to be completely secure–which is a near impossible task. Security can be breached on both hardware and software levels. Modern smartphones often have exotic zero-click exploits, meaning the devices can be hacked without any action by the user of the device. Modern operating systems – be it for desktop or smartphones often have zero-day exploits, meaning the devices running these updated operating systems can be exploited by hackers even on the first day of the update. Exploited devices can act as Byzantine nodes in IoT networks, making the entire network vulnerable and stealing data from other devices. Especially in healthcare systems, the underlying data is extremely sensitive and confidential in nature. Healthcare data not only contain health-related information but also contains identifying information of the patients including social security numbers, social insurance number, and insurance information. Any leak or compromise in such networks might result in catastrophic ramifications for the patient such as identity theft, and customized scamming. Improving the security and privacy of data in healthcare systems is a research challenge.

In the future, cryptosecurity algorithms such as the RSA encryption algorithm, Digital Signature Standard (DSS), and ElGamal can be utilized in all endpoints in IoT networks to provide security. Secure hash algorithms can be utilized to ensure data integrity. Blockchain technology can be leveraged to reduce man-in-the-middle and network eavesdropping attacks [180], [181]. Non-Fungible Token (NFT) technology can save patients from counterfeit products or services. NFTs can also help patients to decentralize their sensitive data and to share healthcare data while keeping their identities anonymous. Biometric authentication techniques such as fingerprints, voice recognition, face recognition, and iris recognition as well as software-based authentication such as two-factor authentication, access cards, and seed phrases should also be standardized in the future for systems connected in IoT networks. Artificial intelligence-based approaches for anomaly detection in networks should be explored in the future for improving security. Due to advances in TinyML [182], it might be possible to run anomaly detection models in the future in resource-constrained devices [183]. These would greatly improve the security and privacy of the users and the transmitted data.

### B. Standardization in IoT

Heterogenous types of devices are interconnected via complicated networks in an IoT environment. Each of these devices might use distinct architectures or protocols for generating or transmitting data. The devices are also different from software and hardware perspectives. The management of such devices including the collaboration of the devices in a software or hardware level, proper addressing, marshalling, de-marshalling, optimization at a protocol or architectural level, and identification is a huge research challenge. Measures should be taken to draft and standardize IoT architecture and network protocols in the future [184].

### C. Data Variability in IoT

Data in Internet of Things can vary in type, size, and formation. Due to the sheer number of connected devices in IoT networks, storing the data in solutions such as in conventional RAID systems or conventional databases is very inefficient. Huge volumes of data are generated in a very short period in IoT networks. Conventional database systems are not capable of storing or efficiently retrieving them. Management of the huge volume of data such as handling spatial or temporal information, handling differing data sizes, and handling different types are massive research challenges. Devices in healthcare systems transmit data in various formats such as 2D image formats, 3D imaging formats, and numerical sensor readings in time-series format.

Big data warehousing and processing frameworks such as Apache Hadoop, Apache ZooKeeper, and Apache Kafka can be utilized to efficiently deal with the huge volume of data in the future [185]–[188]. Some of these frameworks also provide error detection and error recovery of the stored data, which will help in persisting historical data. Some of these frameworks are also capable of dealing with multiple types of data at the same time. These frameworks are also highly cost and space-efficient. Lossless compression models can also be researched to transmit huge volumes of data over IoT networks [189].

### D. Identity in IoT Networks

IoT networks generally consist of a huge number of devices. Every device needs a unique identity in the network. The most used core protocol in standard-based interconnecting networks is IPv4. The IPv4 protocol uses 32 bits to represent unique addresses which provide a maximum of $2^{32}$ address spaces. IPv4 addresses are represented in 4 numbers separated by 4 dots. Due to the huge number of devices connected to IoT networks, it is possible to run out of address space. Healthcare systems have multiple monitoring systems per person. For a sufficiently large elderly population in a single network, the address space for all devices might be insufficient. While a lot of devices support IPv6 nowadays, developing more lightweight and robust solutions for addressing the identified problem is a research challenge.

In the future, IPv6 protocol can be standardized for all IoT networks [190], [191]. IPv6 addresses are represented as eight groups of numbers. Each group contains four hexadecimal digits. The address space of IPv6 is $2^{128}$ which is suitable for addressing the huge number of devices in IoT networks. IPv6 also supports shorthand representation and compression of network addresses, which reduces total used bandwidth over the network.

### E. Lack of Service Description Language

A service description language is a special form of Interface Description Language that is used for describing a particular service. Interface description languages enable programs to interact with one another regardless of origin or language. Similarly, service description language enables services from varying devices and software/hardware combinations to interact with one another. Due to lack of standardization, devices in IoT networks use vastly different protocols, architectures, and data types. The underlying service routines dealing with these differences are also vastly different from one another. A unified service description language would greatly benefit IoT networks by enabling seamless collaboration and interoperability among devices. Creating a unified service description language is a massive research challenge. In the future, service description languages such as XML, and UML. can be extended to create a standardized service description language for IoT networks. Research should also be focused on developing specialized service description language for IoT networks [192], [193].

### F. Design of Service-Oriented Architecture

Service-Oriented architecture is a style of software architecture where components of an application interact with other components of that application or other application via services through utilizing a communication protocol over a network. The connection is independent of vendor or other limiting hardware or software technologies. The design of service-oriented architecture for IoT systems is a huge challenge. On one hand, service-oriented architecture simplifies the management of heterogeneous types of connected devices. On the other hand, service-oriented architecture introduces additional overhead to devices. Most IoT devices are resource-constrained. Adding overhead to these resource-constrained devices slows down the entire network making data processing, collection, and transferring more difficult. Thus, researching lightweight service-oriented architecture with minimum overhead that facilitates big data warehousing and processing is a huge research challenge [194]–[196]. In the future, novel architectures for IoT networks should be developed to be modular and service-oriented in nature.

### G. Hardware Customization

IoT networks currently degenerate into traditional ICT oriented networks. Similarly, the hardware of various intelligent devices is not customized to specific protocols. IoT devices generate huge amounts of data within short amounts of time. Higher bandwidth networks are necessary alongside big data measures to store and process the data in real-time. Developing novel methods to transport the huge amounts of data in commodity networks is a huge research challenge, as it is impossible to deploy high bandwidth high-frequency networks all around us. Similarly, optimizing the hardware of general intelligent things to efficiently use certain universal IoT protocols is a research challenge.

In healthcare systems, various sensors are used to monitor patients. Developing novel hardware design solutions to contain various sensors in small housing would enable a single device to monitor various aspects of a patient. In the future, various bio-inspired optimization techniques such as swarm optimization, and genetic algorithms should be utilized to automatically develop optimized hardware design for various use cases.

### H. Mobility of IoT Devices

In an interconnected world from a network perspective, intelligent devices are expected to disconnect from one network and connect to other networks. In traditional networks, mobility is a huge issue. For mobile nodes, routing tables have to be re-calculated each time nodes move out of range from one another and come into the range of other nodes. This calculation puts a huge overhead on the entire network when numerous nodes with high mobility are present [197]. Modeling and handling highly mobile nodes in IoT networks are a huge research challenge. In healthcare systems, some monitoring devices might be mobile in nature, while others might be stationary. For example, wearable monitoring devices might be highly mobile, while surveillance monitoring devices would be stationary. Patients would wear and bring the highly mobile monitoring devices to their activities of daily life. Thus, optimizing the various aspects of these mobile devices, such as route discovery, routing table generation, efficient data transfer, and reducing data costs is a research challenge.

In the future, artificial intelligence-based approaches such as reinforcement learning-based techniques can be used to develop networks facilitating highly mobile devices [198], [199]. Vehicular Ad-hoc Network (VANET) architectures can be extended in the future to IoT networks involving highly mobile nodes [200].

### I. Reliability in IoT Networks

Reliability is a huge issue in commodity networks. Regardless of network and type of protocol, network packets get lost with the data. While some data warehousing solutions provide checkpointing and state management for error detection and recovery, it is not the norm and requires extensive installation and support. Developing lightweight reliable networks with error detection and error recovery is a huge challenge. Devices connected in IoT networks themselves need to be reliable [201]–[203]. Erroneous reporting might result in faulty directions from central controller servers. Healthcare systems in general require reliable and fast networks. Any drop in data, as well as monitoring results, might be irreversibly harmful for the users. For example, in stroke, diabetes, Alzheimer's, or fall monitoring systems, it is imperative to monitor the patients without interruptions, and any kind of delays in transmission or monitoring might prove fatal for the patients. Developing specialty fault tolerant and state-recovery-capable networks for healthcare systems is a challenge. Similarly, unreliable devices in healthcare IoT networks can result in the loss of life of patients in extreme cases.

In the future, network traffic forecasting models should be developed to predict traffic load and allocate network resources accordingly. Various checkpointing algorithms can be used for error detection and correction in networks. Network designs should also be fault tolerant. Advances in TinyML enable us to deploy and run machine learning models

| Type | Concerns | Recommendations |
|---|---|---|
| Security and Privacy | ▪ Healthcare data is highly sensitive and contains personal identifying information such as social security information, and insurance information. | ▪ More secure overlay networks such as The Onion Router (TOR) network might be used to transfer confidential data.<br>▪ Technologies such as blockchain, and NFT tickets can be used to provide authentic and customized service.<br>▪ Existing security solutions such as RSA, seed phrases, and DSS might be employed in all connection endpoints.<br>▪ Authentication and identity verification methods such as password, fingerprint scanning, signature, voice pattern, and smart card can be employed in application protocols.<br>▪ Artificial intelligence-based approaches can be used to detect anomalies in IoT networks. |
| Standardization and Interoperability | ▪ Different devices in IoT networks use differing application protocols, data representation methods, network protocols, marshalling, de-marshalling, and identification.<br>▪ No standard service description language.<br>▪ Limited interoperability among devices. | ▪ Developing a standardized service description language would benefit the standardization and interoperability of IoT devices.<br>▪ Open-source implementations of vendor-locked applications and networking protocols would enable interoperability.<br>▪ Standardizing application and networking protocols would simplify interoperation among devices in IoT networks. |
| Data Variability and Scalability | ▪ Data transmitted in IoT networks vary by size, type, and representation over network.<br>▪ Massive amounts of data can be transmitted over an IoT network in a very short amount of time.<br>▪ General applications and networks are not scalable in nature. | ▪ Utilizing big data warehousing techniques and open-access application frameworks such as Apache Hadoop, and Apache zookeeper will allow the storing and processing of huge amounts of data.<br>▪ Modularity can be prioritized to make devices in IoT networks more scalable.<br>▪ Lightweight service-oriented architecture of devices would enable scalable big data warehousing and processing.<br>▪ Utilizing available SaaS, PaaS, and DaaS platforms such as AWS, Google App engine would increase the scalability of IoT-orientated solutions.<br>▪ Lossless compression algorithms might be utilized to transmit huge volumes of data over standard IoT networks. |
| Power and Cost Efficiency | ▪ Increasing number of intelligent devices connected in IoT networks has harmful effects on nature.<br>▪ Devices are not as power-efficient as they could be.<br>▪ Setting up and maintaining all-day monitoring via multiple devices is not cost-efficient. | ▪ Solar-powered hardware solutions would increase the sustainability and eco-friendliness of the huge number of devices used in IoT networks.<br>▪ Creating multi-purpose generalized devices would reduce the number of separate devices necessary to monitor or surveil patients or environment. This reduction in number would also reduce the overall cost.<br>▪ Specialized hardware for task-specific purposes might reduce overall power consumption.<br>▪ Optimized software would also reduce the amount of necessary processing, which might optimize power consumption. |
| Reliability and Availability | ▪ Devices used in monitoring purposes in both industrial and healthcare use cases need to be available all the time regardless of network or other failures for avoiding dangerous or life-threatening situations.<br>▪ Intelligent devices used for monitoring and surveillance purposes need to be reliable at all times. Unreliable monitoring can result in erroneous diagnoses which might even lead to death of patients.<br>▪ IoT networks need to maximize availability and reliability. Downtime of networks might lead to loss of valuable data. | ▪ Fault-tolerant development principles should be followed while developing specialized IoT networks.<br>▪ Error detection and recovery techniques in networking such as distributed checkpointing, and coordinated checkpointing can be implemented to increase network reliability.<br>▪ Network traffic analysis and traffic forecasting might be used to anticipate network load, which will help in allocating appropriate resources to increase network availability.<br>▪ Simplified versions of applications must be available in devices to facilitate basic services even during network failure or downtime.<br>▪ TinyML deployment techniques can be leveraged to ensure minimal functionality during network downtime.<br>▪ Sensing modules in devices must be within standardized error tolerance limits and must undergo scheduled testing to guarantee utmost reliability.<br>▪ Using state-of-the-art battery technologies might improve the reliability and availability of devices. |
| Identity and Mobility | ▪ IPv4 network protocol might not be sufficient to address millions of devices connected in IoT networks.<br>▪ Networks and routing protocols are not efficient in dealing with numerous highly mobile devices in IoT networks. | ▪ IPv6 can be adopted as the standard Internet protocol for IoT networks.<br>▪ Artificial intelligence approaches such as reinforcement learning techniques can be utilized to develop and test new routing protocols capable of efficiently dealing with highly mobile nodes.<br>▪ Vehicular Ad-hoc Networks already provide networking solutions for mobile and stationary nodes in networks. Relevant components from VANETs can be evolved for use in IoT networks. |

in resource-constrained environments. In the future, devices connected in IoT networks should use TinyML to run basic models to provide basic service functionalities during network downtime or other failures. This would increase the reliability factor of the devices. To reduce the downtime of the devices, advanced battery solutions should be used in the future.

## J. Scalability in IoT Networks

IoT networks might have millions or billions of interconnected nodes. General networks and services are not scalable in nature. Scalability is a huge challenge for such IoT networks [204], [205]. In healthcare systems, each patient might have multiple types of monitoring going on at the same time. For example, elderly patients might be monitored for heart diseases, glucose levels, diabetes, Parkinson's, Alzheimer's, and falls at the same time. For a sufficiently large elderly population, the number of these systems would reach millions, requiring specialty scalable services and applications.

In the future, the potential of using existing Database-as-a-Service (DaaS), Platform-as-a-Service (PaaS), and Software-as-a-Service (SaaS) to increase the scalability of IoT services and networks should be explored. The research focus should be on developing modular IoT solutions that can scale on need.

### K. Availability in IoT Networks

IoT has applications in industrial and healthcare domains among other potential application areas. Industrial devices are placed for continuous monitoring of various parameters such as temperature, humidity levels, and pressure. Even during network failures, the devices should be able to provide basic monitoring services to avoid catastrophic failures in industrial cases. In the healthcare space, monitoring devices need to be available at all times similar to industrial use cases. Lack of availability might result in serious bodily harm or even death. Devices in IoT networks must provide basic monitoring and action services even when they are disconnected from networks or when network failures happen. The availability should be implemented in both hardware and software levels.

Similar to the reliability, research should be focused on deploying basic models into devices using TinyML, which would increase overall availability. Novel error detection and error correction in fault tolerant networks should be explored to increase the availability of the network.

### L. Interoperability in IoT Networks

Interoperable applications and services can communicate and transmit data among themselves regardless of hardware or software differences. However, interoperability is very hard to achieve. New protocols or architectures that are constantly being developed do not always support all legacy protocols and architectures, making data transmission and interaction between devices very complex and problematic. Similarly, operating systems and various applications also have different implementations, making data sharing very difficult. Vendors often-time have patented and exclusive implementations that make their devices only compatible with other devices of the same vendor. Thus, interoperability is a huge research challenge for IoT [206], [207]. Interoperability is very important in healthcare systems as various medical equipment have differing hardware, protocol implementations, and data representation. For a smooth deployment and operation of IoT networks in the healthcare sector, the various devices and services need to be interoperable.

In the future, the research focus should be on developing open-access network architecture and protocols to facilitate interoperability among devices. Open-access implementations of specialty architecture and protocols would also facilitate this issue.

### M. Power and Cost Efficiency in IoT Networks

IoT networks have millions of nodes. Power and cost are huge concerns for such a huge number of devices. The entire world is also experiencing various global natural catastrophes due to global warming and climate change caused by extracting and depending on fossil fuels, trashing the oceans with electronic waste, and chemical waste. With an increased understanding of the causes behind the sudden climate change and global warming, researchers are increasingly moving towards sustainable, eco-friendly, and bio-degradable solutions. Making IoT devices more eco-friendly, cost-efficient, and power-efficient is a massive research challenge [208]. Similarly, converting medical intelligent monitoring devices in healthcare IoT networks into more multi-purpose, power and cost-efficient, long-lasting, eco-friendly versions is a research challenge.

In the future, research should be focused on simplifying and combining multiple sensing modalities in single devices to reduce the number of total sensors necessary to monitor and surveil users. The focus should also be on developing more area-surveillance solutions. Area-surveillance technologies would reduce the number of devices per individual. Reducing the total number of devices needed per user would also drastically reduce the cost to effectively deploy an IoT solution. In the future, research should be focused on transforming commodity devices into solar-powered versions. Research on hardware and software optimization would also increase the power the efficiency of devices. Making the devices solar-powered would reduce the overall pressure on fossil fuel, which has beneficial effects on nature.

The summary of challenges in large-scale deployment of IoT networks and future research recommendations are shown in Table IV to provide a broad idea to the readers in a short look.

## VIII. CONCLUSION

The Internet has significantly shaped the way people live, allowing individuals to interact with one another on a virtual level in a variety of situations ranging from professional to social. The Internet of Things could contribute meaningfully to this system by enabling communication among connected devices, which tends to result in the dream of "anymedia, anytime, anything, anywhere" connectivity. This paper focuses on the cutting-edge IoT device capabilities, architectures, and protocols that describe the IoT conceptual framework. The IoT device capabilities for hardware and software tools are thoroughly demonstrated in this paper. There have been discussions about possible IoT architectures, such as the conventional three-layer architecture, the SoA based architecture, and the middleware-based architecture. Additionally, the IoT communications protocols with their distinguished features are incorporated briefly in this paper. We have highlighted various IoT applications for smart healthcare that have been developed in the recent decades to ensure some facilities for medical practitioners such as monitoring and identifying a variety of health problems, measuring various health factors, and offering diagnosis and treatment at remote locations. Lastly, the current challenges and gaps related to the development of IoT-based healthcare applications, along with future improvement and research focus in the coming years have been explored here. Hopefully, this paper will provide a stable platform for scholars and scientists willing to engage in learning more about IoT technologies in the context to fully grasp the general architecture and roles of the multiple elements and processes that comprise the concept of Internet of Things.

# Appendix

## Table A
### List of Abbreviations

| Abbreviation | Definition | Terminology |
|---|---|---|
| ADC | Analog to Digital Converter | General |
| AI | Artificial Intelligence | General |
| AMQP | Advanced Message Queuing Protocol | Technical |
| AWS | Amazon Web Services | General |
| BLE | Bluetooth Low Energy | Technical |
| BP | Blood Pressure | Medical |
| CoAP | Constrained Application Protocol | Technical |
| CoRE | Constrained RESTful Environments | Technical |
| CPU | Central Processing Unit | General |
| DaaS | Database-as-a-service | Technical |
| DDS | Data Distribution Service | Technical |
| DSS | Digital Signature Standard | Technical |
| ECG | Electrocardiogram | Medical |
| GPIO | General Purpose Input Output | Technical |
| GPS | Global Positioning System | General |
| HTTP | Hypertext Transfer Protocol | Technical |
| IETF | Internet Engineer Task Force | General |
| I/O | Input/Output | General |
| IPv6 | Internet Protocol Version 6 | Technical |
| ISO/IEC | International Organization for Standardization/International Electrotechnical Commission | General |
| IoT | Internet of Things | Technical |
| LoWPAN | Low-Power Wireless Personal Area Networks | Technical |
| ML | Machine Learning | General |
| M2M | Machine to Machine | Technical |
| MQTT | Message Queue Telemetry Transport | Technical |
| NFT | Non-Fungible Token | Technical |

Intelligence: Multiaccess Edge Computing for 5G and Internet of Things," *IEEE Internet Things J.*, vol. 7, no. 8, pp. 6722-6747, Aug. 2020.

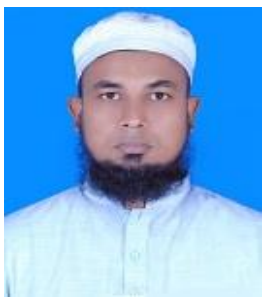

**Md. Milon Islam** is currently pursuing his PhD at the Centre for Pattern Analysis and Machine Intelligence in the Department of Electrical and Computer Engineering at the University of Waterloo, Canada. He received his B.Sc. and M.Sc. degree in Computer Science and Engineering (CSE) from the Khulna University of Engineering & Technology, Khulna, Bangladesh, in 2016 and 2019 respectively, where he is currently working as an Assistant Professor (on leave) in CSE. In 2017, he joined the Department of Computer Science and Engineering, Khulna University of Engineering & Technology, as a Lecturer. He has published several numbers of research papers in peer reviewed journals, book chapters and conferences. In addition, he is reviewer of several reputed journals and conferences. His research interests include machine learning and its application, deep learning, intelligent systems design, health informatics, Internet of Things (IoT), and to solve real life problems with the concept of computer science.

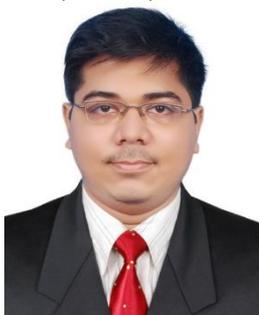

**Sheikh Nooruddin** is currently pursuing his Master of Applied Science (MASc) degree in Electrical and Computer Engineering (ECE) with a specialization in Pattern Analysis and Machine Intelligence (PAMI) at the University of Waterloo, Ontario, Canada. He received his B.Sc. Eng. in Computer Science and Engineering degree from Khulna University of Engineering & Technology, Bangladesh in 2020.
His research interests include computer vision, artificial intelligence, human activity recognition, Internet of things, and medical signal processing. He has authored and co-authored multiple journals and conference articles published by reputed peer-reviewed international publishers including IEEE, Elsevier, and Springer.

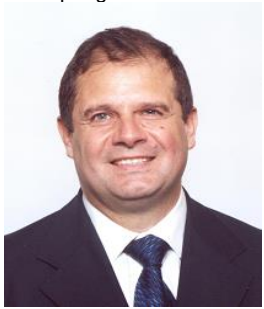

**Fakhri Karray** is a Professor and Provost of the Mohamed Bin Zayed University of AI in the UAE. He has served as the University Research Chair Professor in Electrical and Computer Engineering and the founding co-director of the Institute of Artificial Intelligence at the University of Waterloo. He holds the Loblaw's Research Chair in Artificial Intelligence. Dr. Karray's research work spans the areas of intelligent systems and operational artificial intelligence as applied to smart devices and man machine interaction systems through speech, gesture, and natural language. He has authored extensively in these areas and has disseminated his work in journals, conference proceedings, and textbooks. He is the co-author of two dozen US patents, has chaired/co-chaired several international conferences in his area of expertise and has served as keynote/plenary speaker on numerous occasions. He has served as the associate editor/guest editor for a variety of leading journals in the field, including the IEEE Transactions on Cybernetics, the IEEE Transactions on Neural Networks and Learning Systems, the IEEE Transactions on Mechatronics, the IEEE Computational Intelligence Magazine. He has served as the University of Waterloo's Academic Advisor for Amazon's Alexa Fund Fellowship Program and is a Fellow of the Canadian Academy of Engineering, a Fellow of the Engineering Institute of Canada and a Fellow of the IEEE.

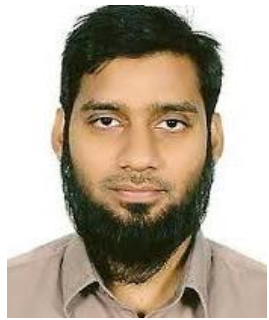

**Ghulam Muhammad** (Senior Member, IEEE) is a professor in the Department of Computer Engineering, College of Computer and Information Sciences at King Saud University (KSU), Riyadh, Saudi Arabia. Prof. Ghulam received his Ph.D. degree in Electronic and Information Engineering from Toyohashi University and Technology, Japan in 2006, M.S. degree from the same university in 2003. He received his B.S. degree in Computer Science and Engineering from Bangladesh University of Engineering and Technology in 1997. He was a recipient of the Japan Society for Promotion and Science (JSPS) fellowship from the Ministry of Education, Culture, Sports, Science and Technology, Japan. His research interests include signal processing, machine learning, IoTs, medical signal and image analysis, AI, and biometrics. Prof. Ghulam has authored and co-authored more than 300 publications including IEEE / ACM / Springer / Elsevier journals, and flagship conference papers. He owns two U.S. patents. He received the best faculty award of Computer Engineering department at KSU during 2014-2015. He has supervised more than 15 Ph.D. and Master Theses.